\documentclass[12pt]{article}

\usepackage{scicite}
\usepackage{graphicx}
\usepackage{wrapfig}
\usepackage[hyphens]{url}
\usepackage{times}
\usepackage{tabu}
\usepackage{longtable}
\usepackage{amsmath}

\newenvironment{sciabstract}{%
\begin{quote} \bf}
{\end{quote}}

\title{ALBATROSS: Cheap Filtration Based Geometry via Stochastic Sub-Sampling}

\author
{Andrew J. Stier$^{1}$$^\ast$, Naichen Shi$^{2,3}$,\\
Raed Al Kontar$^{4}$, Chad Giusti$^{5}$, 
 Marc G. Berman$^{6}$\\
\\
\normalsize{$^{1}$The Santa Fe Institute,}\\
\normalsize{$^{2}$Department of Industrial Engineering and Management Sciences, Northwestern University.}\\
\normalsize{$^{3}$Department of Mechanical Engineering, Northwestern University.}\\
\normalsize{$^{4}$Industrial and Operations Engineering Department, University of Michigan}\\
\normalsize{$^{5}$Department of Mathematics, Oregon State University}\\
\normalsize{$^{6}$Psychology Department, The University of Chicago}\\
\\
\normalsize{$^\ast$To whom correspondence should be addressed; E-mail:}\\
\normalsize{andrewstier@uchicago.edu}
}

\date{}

\begin{document} 


\baselineskip24pt


\maketitle

\begin{sciabstract}
 Topological data analysis (TDA) detects geometric structure in biological data. However, many TDA algorithms are memory intensive and impractical for massive datasets. Here, we introduce a statistical protocol that reduces TDA's memory requirements and gives access to scientists with modest computing resources. We validate this protocol against two empirical datasets, showing that it replicates previous findings with much lower memory requirements. Finally, we demonstrate the power of the protocol by mapping the topology of functional correlations for the human cortex at high spatial resolution, something that was previously infeasible without this novel approach.
\end{sciabstract}

\begin{figure}[hbpt!]
\centering
\includegraphics[width=.7\textwidth]{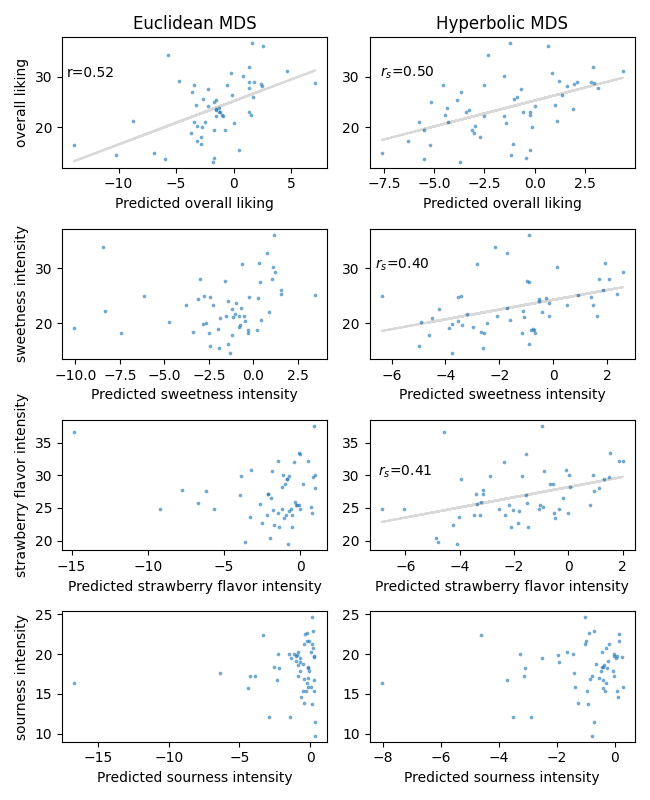}
\caption{Euclidean multi dimensional scaling fails to capture the perceptual axes of strawberry odor space. We performed Euclidean and Hyperbolic multidimensional scaling on a dataset of monomolecular odor concentrations in samples of strawberries. Although the embedding was slightly more representative of the data in the Euclidean case (see Supplementary Figure \ref{fig:mdsCompStats}), attempts to find perceptual axes (see Methods) in monomolecular odor space were successful only for overall liking in Euclidean space, but were successful for 3 of the 4 perceptual ratings in Hyperbolic space. Further, a linear axis is a significantly better fit for all perceptual dimensions in Hyperbolic space while the sole linear axis found in Euclidean space is poorly fit with residual plots suggestive of remaining non-linearities in the data (Supplementary Figure \ref{fig:mdsCompStats}). Predicted values were computed via leave-one-out cross validation.}
\label{fig:mdsComp}
\end{figure}

The premise of topological data analysis~\cite{sizemore2019importance,wasserman2018topological} is that understanding the topological structure of biological data is critical for deciding which analysis techniques to employ and guiding which types of mechanistic theories to adopt. This is because reasoning about biological mechanisms is often carried out in the low-dimensional spaces resulting from dimensionality reduction techniques when large amounts of data are available~\cite{shine2019human,nair2023approximate,suresh2020neural}. Common algorithms that are employed in the early stages of data pre-processing include, t-distributed stochastic neighbor embedding (t-SNE)~\cite{hinton2002stochastic}, multidimensional scaling (MDS)~\cite{mead1992review}, and eignenvalue decomposition (see~\cite{dahmen2019second} for an example from neuroscience).   However, most of these algorithms default to the assumption that data are representative of a flat and linear Euclidean space. This assumption frequently does not hold for biological data: among other examples~\cite{giusti_2018,giusti_clique_2015}, the molecular composition of odors~\cite{zhou2018hyperbolic}, patterns of gene expression~\cite{zhou2021hyperbolic}, the activity of neurons in visual cortex~\cite{singh_topological_2008} as well as aspects of human visual perception of space~\cite{luneburg1947mathematical,indow2004global} are more appropriately represented by spaces with negative and positive curvature (i.e., non-Euclidean spaces). Thus, standard dimensionality reduction techniques may not appropriately capture the underlying structure of the data and may result in inaccurate inferences about biological mechanisms (see Figure \ref{fig:mdsComp} \& Supplementary Figure \ref{fig:mdsCompStats} for an example). As a result, topological data analysis (TDA), which allows for the detection of inherent geometric structure in data, has become an increasingly popular tool to analyze publicly available biological datasets. However, TDAs applicability has been limited because state-of-the-art algorithms have memory requirements that scale exponentially with the amount of data~\cite{bauer2021ripser,henselman2016matroid}. Here we propose a statistical protocol \textbf{ALBATROSS} which circumvents these memory requirements and allows TDA to be run on large datasets.

In neuroscience, where measurements often focus on the co-occurrence between events or processes~\cite{giusti_clique_2015}, TDA is commonly applied to direct and indirect measurements of neural activity and to measurements and perceptual ratings of stimuli that invoke such activity. Typically, these data are transformed into temporal correlations or used to measure the similarity between stimuli~\cite{curto2017can,giusti_twos_2016,zhou2018hyperbolic}. Specifically, such data are summarized in square adjacency matrices, $A_{ij}$, which encode pairwise similarity between units of interest (e.g., neurons, brain regions, stimuli, participants, etc.) in a continuous nature.  When adjacency matrices are analyzed with tools from network science~\cite{bassett2017network,sporns2010networks}, the construction of networks from $A_{ij}$ requires setting a portion of the entries to zero to obtain a network that is sparse. This problematically discards most of the information contained in the magnitudes of $A_{ij}$, which are typically biologically relevant~\cite{giusti_twos_2016}. In contrast, TDA makes full use of this information by analyzing filtered simplicial complexes which are combinatorially formed from the cliques (or all-to-all connected subgraphs) of all possible discrete network constructions from $A_{ij}$ (also know as the clique complex of the weighted network, $A_{ij}$, see Methods). TDA based on filtered simplicial complexes has been used to infer the geometric properties of evoked and spontaneous population activity in the visual cortex~\cite{singh_topological_2008}, of perceptual and molecular properties of odors~\cite{zhou2018hyperbolic}, and of differences in geometric properties between brain artery trees among younger and older humans~\cite{bendich_persistent_2016}. 

\begin{figure}[hbpt!]
\centering
\includegraphics[width=1\textwidth]{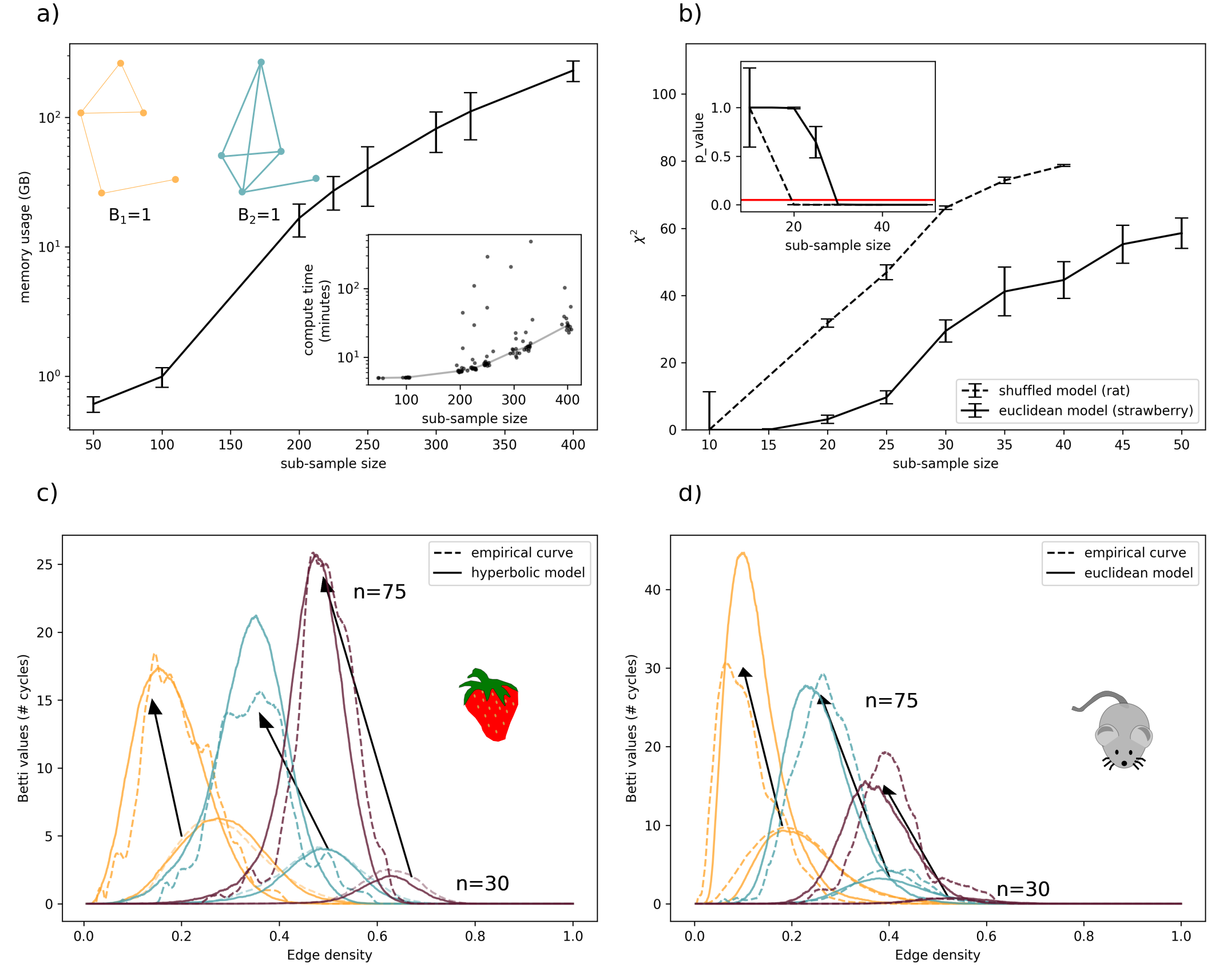}
\caption{a) Computation of Betti curves requires exponentially more memory with increasing dataset size. Inset: Compute time requirements also increase with size, though more slowly. Top Left: Cartoon showing example Betti numbers for two networks. b) $\chi^2$ test statistics for the null models increase with sub-sample size when using ALBATROSS. Statistical inference of global geometric structure is reliable for $n>30$. Note that significant p-values and higher $\chi^2$ statistics indicate that the empirical data \textit{do not} fit the candidate geometric space. Inset: p-values decrease with $n$. c,d) ALBATROSS finds well fit geometries for most values of $n$. Note that Betti numbers increase with sub-sample size. This is true for a dataset of strawberry odors, c), and a dataset of rat hippocampal place cell firing, d). Each betti curve is plotted in a different color $\beta_1$, $\beta_2$, $\beta_3$ in yellow, teal, and purple, respectively. Arrows indicate how the curves change going from low values of $n$ to high values of $n$.}
\label{fig:one}
\end{figure}

However, the computation of filtered simplicial complexes is computationally difficult and expensive~\cite{giusti_clique_2015, giusti_twos_2016} - at worst, memory requirements grow exponentially, O(2$^N$) with the size of $A_{ij}$. With state of the art software programs~\cite{henselman2016matroid,bauer2021ripser} which perform close to theoretical limits, our computational experiments (Figure \ref{fig:one}) and others~\cite{bauer2021ripser,henselman2016matroid,malott2020topology} demonstrate that in practice the memory requirements are much better than the worst case scenario. Still, we find that computing filtered simplicial complexes matrices $A_{ij}$ with 300 rows and columns requires a median of 71 GB of memory (Figure \ref{fig:one}a) and is computationally impractical for more than 400 rows and columns (Supplementary Table \ref{tab:memory}). Thus, TDA has remained inaccessible to researchers without large institutional computing resources and is impractical for large datasets even when such resources are available.  This is unfortunate because TDA techniques promise to help identify structure in massive biological datasets and are backed by close to 100 years of theoretical groundwork in algebraic topology  which can help scientists interpret discovered topological structure, inform follow-up analyses, and guide the development of new theoretical models~\cite{sizemore2019importance}.

Here, we introduce a statistical protocol to analyze filtered simplicial complexes. This protocol decreases the memory and computational barriers for TDA of large-scale datasets. This protocol is based on a number of recent mathematical results~\cite{solomon2021geometry,malott2020topology} which demonstrate that measures on filtered simplicial complexes of $A_{ij}$ can be reliably and stably estimated by various forms of stochastic sub-sampling of $A_{ij}$ and subsequently computing filtered simplicial complexes on these much smaller sub-samples. This protocol, che\textbf{A}p fi\textbf{L}tration \textbf{BA}sed geome\textbf{tr}y via st\textbf{O}chastic \textbf{S}ub-\textbf{S}ampling: \textbf{ALBATROSS} (\url{https://github.com/enlberman/albatross}), massively reduces the memory requirements for filtered-simplicial-complex-based TDA measurements to the point that basic TDA analyses of large datasets can be run locally on moderately equipped personal computers (Supplementary Table \ref{tab:pc}). 

Beyond ALBATROSS's stochastic subsampling, finding methods of data reduction that preserve topological properties sufficiently is an active area of research. Recent proposals have used various forms of clustering on point clouds to reduce the memory and computational requirements in ways that can outperform stochastic subsampling~\cite{koyama2024distilled,koyama2023faster,malott2020topology}. However, these techniques have yet to be studied for the clique complexes of weighted networks, and the analyses presented here show that stochastic subsampling performs well outside of the context of point clouds in which it has been previously studied~\cite{solomon2021geometry}.

We focus here on applications of the ALBATROSS protocol to statistically infer the global geometry of biological data.\footnote{Though we note that since classic network science~\cite{sporns2010networks,bassett2017network} analyses are a subset of TDA, in that they consider a small subset of a full filtered simplicial complex, it is possible to apply ALBATROSS's statistical protocols to compute traditional network measures such as identifying the presence of rich clubs~\cite{giusti_twos_2016}.} Specifically, we compute the first $n$ Betti numbers~\cite{giusti_clique_2015}, which count the number of $n$-dimensional cycles -- or holes (Figure \ref{fig:one}a, top left) -- in a thresholded and binarized network. These Betti numbers, can be computed on the entire filtered simplicial complex to generate the n$^{th}$ Betti curve~\cite{solomon2021geometry,giusti_twos_2016} (Figure \ref{fig:one} c\&d). The shape and area under these curves can be indicative of global geometric structure~\cite{giusti_clique_2015,zhou2018hyperbolic,giusti_twos_2016,zhang2023hippocampal,bubenik2020persistent}, if such structure exists. In practice, the first three Betti curves are typically sufficient to infer global geometry~\cite{giusti_clique_2015,zhou2018hyperbolic,giusti_twos_2016,zhang2023hippocampal,bubenik2020persistent}. For example, computing the first three Betti curves on small datasets has demonstrated that the activity of neurons in visual cortex is consistent with sampling from the surface of a sphere~\cite{singh_topological_2008}, that activity from hippocampal place field neurons is consistent with sampling from a multi-dimensional hyperbolic space(i.e., a space with negative curvature which is a continuous representation of a hierarchical organization)~\cite{zhang2023hippocampal}, and that the molecular makeup of odor stimuli and human perceptions of those odors is consistent with sampling from a hyperbolic shell ~\cite{zhou2018hyperbolic}. These studies have shown that inferring global topological geometry provides critical insight into patterns of neural activity~\cite{singh_topological_2008,giusti_clique_2015,zhang2023hippocampal} and provides insight into whether neural systems take advantage of inherent structure in natural stimuli~\cite{zhou2018hyperbolic}.

It is important to recognize that in all of these examples, Betti curves alone are insufficient to conclude that the data come from some underlying geometry. Instead concluding, e.g., ~\cite{zhou2018hyperbolic} that olfaction space is hyperbolic requires 1) statistical consistency of empirical Betti curves with Betti curves from a hyperbolic model space, 2) theoretical motivation that the underlying space is hyperbolic (mono-molecular odors are produced by hierarchies of cellular processes) and, 3) a demonstration that the inferred geometry provides some practical explanatory or predictive advantage (e.g., that hyperbolic embeddings explain more dimensions of human perceptions of odors, see Figures \ref{fig:mdsComp} \& Supplementary Figure \ref{fig:mdsCompStats}).

\subsection*{The ALBATROSS protocol}
The ALBATROSS protocol focuses on only the first point above and allows researchers to check the statistical consistency of candidate geometries with empirical data. It has three steps: (1) the computation of empirical Betti curves, (2) a search for the best-fit geometric spaces, and, (3) statistical inference.  Step (1) employs stochastic sub-sampling to compute average empirical Betti curves, $\beta_{emp}$ from experimental data. Step (2) consists of a search to find the best-fit geometries by comparing $\beta_{emp}$ to the average Betti curves for each candidate geometry (see Online Methods). Finally, step (3), takes advantage of the central limit theorem (CLT) to produce a test statistic for the fit between geometric models and empirical data (see Online Methods). In this step, large test statistics and significant p-values indicate a bad fit between the empirical data and candidate spaces (Figure \ref{fig:one}b).

The first two steps both rely on stoachstic sub-sampling. Stochastic sub-sampling involves selecting a sub-sample size, $n$, and the number of iterations, $I$. For each iteration, a random sub-sample of $A_{ij}$ with size $n$ is chosen and the resulting square matrix is used to compute the first three Betti curves (via the Eirene~\cite{henselman2016matroid} software package), which are averaged across all iterations. These average Betti curves are expected to be representative of the geometry of the full adjacency matrix~\cite{solomon2021geometry}. To evaluate candidate geometric spaces, the same procedure is applied to $A_{ij}^{candidate}$, which is the adjacency matrix constructed by computing the distances between points randomly sampled from the candidate geometric space. This protocol computes Betti curves with a constant amount of time and memory usage that is only dependent on the subsample size, $n$ (Figure \ref{fig:one}).

We applied this protocol to previously analyzed datasets of strawberry odors~\cite{zhou2018hyperbolic} and rat hippocampal place cell firing~\cite{giusti_clique_2015,zhang2023hippocampal}. ALBATROSS was able to reproduce the previous findings in these datasets, rejecting euclidean geometry in favor of hyperbolic geometry for the strawberry odors and rejecting a shuffled model in favor of a geometric model for the hippocampal place cells (Figure \ref{fig:one}).  ALBATROSS is able to find well fitting geometries for these empirical datasets with small sub-sample sizes (Supplementary Figure \ref{fig:smallmodels}). However, because the number of possible cycles (and hence the maximum possible Betti number) decreases with the size of the adjacency matrix (Figure \ref{fig:one} c\&d), statistical inference is only reliable above a sub-sample size of $n=30$ with at least $I=100$ iterations (Supplementary Figure \ref{fig:iters}). \begin{figure}[hbpt!]
\centering
\includegraphics[width=1\textwidth]{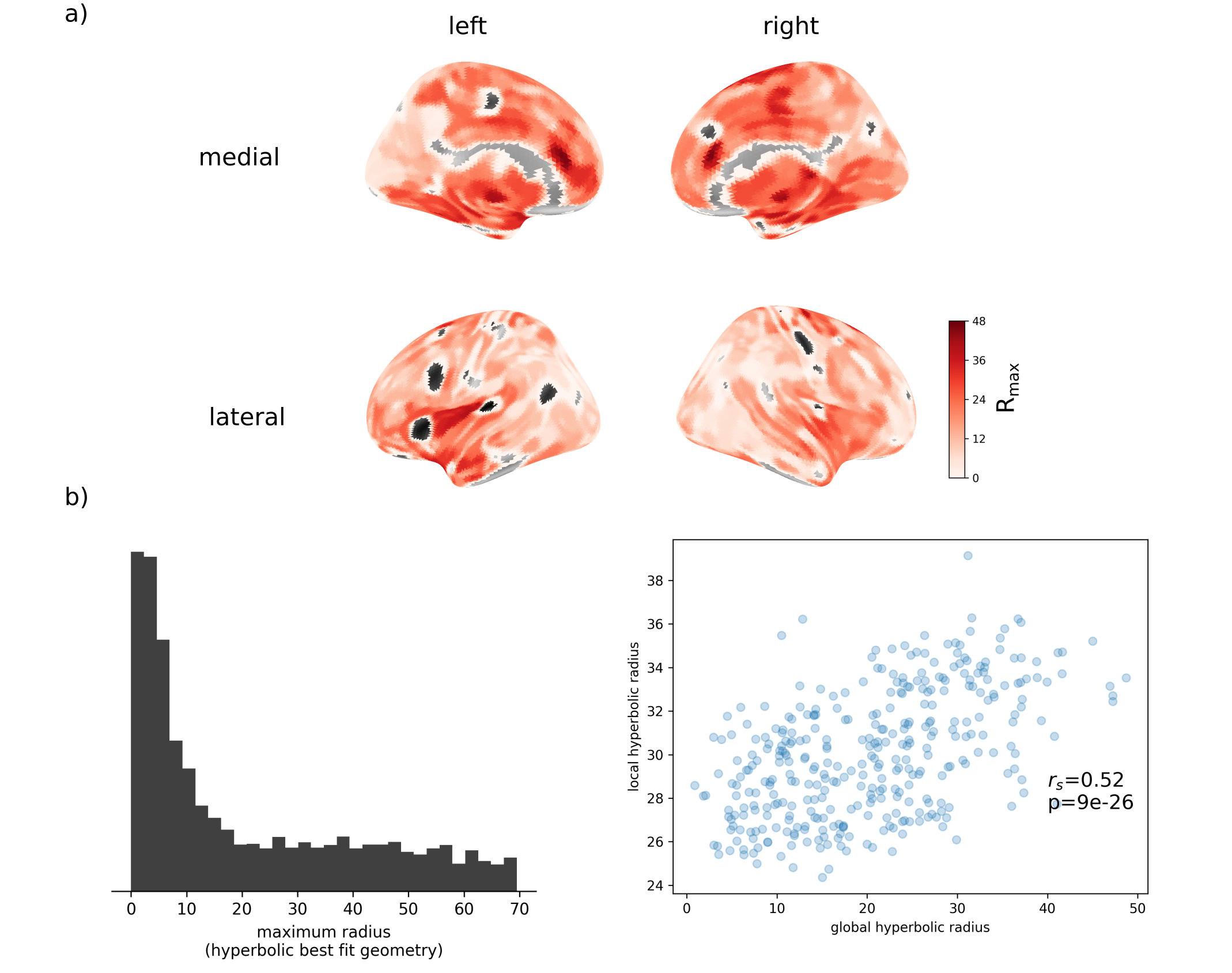}
\caption{Topology of functional correlations across human
cortex. a) ALBATROSS was run for voxel-wise resting state fMRI functional connectivity (rsFC) for 392 previously defined functional parcels. All parcels had topology consistent with a hyperbolic shell. b) Most brain regions have small hyperbolic radii $r<15$, with an approximate uniform distribution at larger radii. c) ALBATROSS was also run on the 392x392 rsFC matrix of the mean time-courses. These data fit a hyperbolic model with a maximum radius of 14. Multidimensional scaling revealed correspondence (Spearman's $r_s = 0.52; p<0.001$) between the radial coordinate of the parcels in 3-D hyperbolic space and the local geometry of their voxelwise rsFC matrices. }
\label{fig:two}
\end{figure} While the specific best-fit model parameters varied slightly depending on $I$ and $n$, the general features of inferred geometric spaces were consistent across these parameter choices and with previous analyses of these data (Supplementary Tables \ref{tab:modelparmsStraw}\&\ref{tab:modelparmsHippo}).

\subsection*{ALBATROSS in a high resolution neuroimaging dataset}

Finally, to demonstrate the power of this protocol in analyzing massive datasets, we mapped local and global geometric structure across the human brain from functional magnetic resonance imaging (fMRI) resting state functional connectivity  matrices(rsFC) in 10 individuals (see Online Methods, Figure \ref{fig:two}). fMRI involves measuring the change in blood flow over time in approximately 100,000 small brain regions and functional connectivity is computed by correlating the fMRI time courses of two regions. With ALBATROSS, we inferred the topology of for 392 previously defined regions from the rsFC matrices of their constituent sub-regions~\cite {craddock2012whole} and for the 392x392 matrix of the regions' mean time-courses. Since some of these parcels contained up to 600 sub-regions, this analysis is not possible without ALBATROSS. 

Surprisingly, the global 392x392 matrices of all ten individuals are consistent with Hyperbolic geometry while only the matrices for four individuals are also consistent with Euclidean geometry.  Across all ten individuals, 353 of the 392 brain regions have local geometry consistent with hyperbolic geometry, while 166 brain regions are consistent with euclidean geometry and 0 are consistent with an a-geometric shuffled model~\cite{giusti_2018,zhou2021hyperbolic,zhang2023hippocampal}. Only 39 of the 392 brain regions are inconsistent with any of the three candidate spaces (hyperbolic, euclidean, or shuffled). The majority of brain regions have hyperbolic best-fit spaces with low radii with $r<15$ (Figure \ref{fig:two}b), and there is an approximately uniform distribution of regions with larger hyperbolic radii. Regions with large $r$ tend to be more medially located, while regions with low $r$ tend to be located more laterally and in sensory areas (Figure \ref{fig:two}a). Importantly, these patterns are not the result of spatially varying noise levels (Supplementary Figure \ref{fig:noise}).

Additionally, the radial coordinates of each region (obtained via hyperbolic multi-dimensional scaling, see Online Methods) are significantly correlated with the maximum radius, $r$, of their local best-fit geometry (Figure \ref{fig:two}c, $r_s=0.52;\ p<0.001$). This correspondence between global radial coordinates and the properties of local topological spaces suggests that local behavior may be driven, in part, by the global functional organization of the brain, something which has been suggested previously~\cite{jiang2021fundamental}. Follow-up analyses may help quantify the hierarchical relationships between different parts of the brain and provide greater insight into the brain's functional organization.

In summary, ALBATROSS reduces the memory requirements of TDA which will allow researchers without extensive computational resources to employ these techniques in their analyses and in large biological datasets. We hope this will speed the continued proliferation of TDA techniques, contribute to the discovery of inherent structure in recently available massive biological datasets, and facilitate more accurate analyses of biological data.

\bibliography{scibib}

\begin{thebibliography}{10}

\bibitem{sizemore2019importance}
A.~E. Sizemore, J.~E. Phillips-Cremins, R.~Ghrist, D.~S. Bassett, {\it Network
  Neuroscience\/} {\bf 3}, 656 (2019).

\bibitem{wasserman2018topological}
L.~Wasserman, {\it Annual review of statistics and its application\/} {\bf 5},
  501 (2018).

\bibitem{shine2019human}
J.~M. Shine, {\it et~al.\/}, {\it Nature neuroscience\/} {\bf 22}, 289 (2019).

\bibitem{nair2023approximate}
A.~Nair, {\it et~al.\/}, {\it Cell\/} {\bf 186}, 178 (2023).

\bibitem{suresh2020neural}
A.~K. Suresh, {\it et~al.\/}, {\it elife\/} {\bf 9}, e58848 (2020).

\bibitem{hinton2002stochastic}
G.~Hinton, S.~T. Roweis, {\it NIPS\/} (Citeseer, 2002), vol.~15, pp. 833--840.

\bibitem{mead1992review}
A.~Mead, {\it Journal of the Royal Statistical Society: Series D (The
  Statistician)\/} {\bf 41}, 27 (1992).

\bibitem{dahmen2019second}
D.~Dahmen, S.~Gr{\"u}n, M.~Diesmann, M.~Helias, {\it Proceedings of the
  National Academy of Sciences\/} {\bf 116}, 13051 (2019).

\bibitem{giusti_2018}
C.~Giusti, Algebraic topology and neuroscience: A bibliography (2018).

\bibitem{giusti_clique_2015}
C.~Giusti, E.~Pastalkova, C.~Curto, V.~Itskov, {\it Proceedings of the National
  Academy of Sciences\/} {\bf 112}, 13455 (2015).

\bibitem{zhou2018hyperbolic}
Y.~Zhou, B.~H. Smith, T.~O. Sharpee, {\it Science advances\/} {\bf 4}, eaaq1458
  (2018).

\bibitem{zhou2021hyperbolic}
Y.~Zhou, T.~O. Sharpee, {\it Iscience\/} {\bf 24}, 102225 (2021).

\bibitem{singh_topological_2008}
G.~Singh, {\it et~al.\/}, {\it Journal of Vision\/} {\bf 8}, 11 (2008).

\bibitem{luneburg1947mathematical}
R.~K. Luneburg  (1947).

\bibitem{indow2004global}
T.~Indow, {\it The global structure of visual space\/}, vol.~1 (World
  Scientific, 2004).

\bibitem{bauer2021ripser}
U.~Bauer, {\it Journal of Applied and Computational Topology\/} {\bf 5}, 391
  (2021).

\bibitem{henselman2016matroid}
G.~Henselman, R.~Ghrist, {\it arXiv preprint arXiv:1606.00199\/}  (2016).

\bibitem{curto2017can}
C.~Curto, {\it Bulletin of the American Mathematical Society\/} {\bf 54}, 63
  (2017).

\bibitem{giusti_twos_2016}
C.~Giusti, R.~Ghrist, D.~S. Bassett, {\it Journal of Computational
  Neuroscience\/} {\bf 41}, 1 (2016).

\bibitem{bassett2017network}
D.~S. Bassett, O.~Sporns, {\it Nature neuroscience\/} {\bf 20}, 353 (2017).

\bibitem{sporns2010networks}
O.~Sporns, {\it Networks of the Brain\/} (MIT press, 2010).

\bibitem{bendich_persistent_2016}
P.~Bendich, J.~S. Marron, E.~Miller, A.~Pieloch, S.~Skwerer, {\it The Annals of
  Applied Statistics\/} {\bf 10} (2016).

\bibitem{malott2020topology}
N.~O. Malott, A.~M. Sens, P.~A. Wilsey, {\it 2020 IEEE International Conference
  on Big Data (Big Data)\/} (IEEE, 2020), pp. 2681--2690.

\bibitem{solomon2021geometry}
E.~Solomon, A.~Wagner, P.~Bendich, {\it arXiv preprint arXiv:2101.12288\/}
  (2021).

\bibitem{koyama2024distilled}
M.~A. Koyama, V.~Robins, K.~Turner, {\it arXiv preprint arXiv:2412.07805\/}
  (2024).

\bibitem{koyama2023faster}
M.~A. Koyama, F.~Memoli, V.~Robins, K.~Turner, {\it arXiv preprint
  arXiv:2307.16333\/}  (2023).

\bibitem{zhang2023hippocampal}
H.~Zhang, P.~D. Rich, A.~K. Lee, T.~O. Sharpee, {\it Nature Neuroscience\/}
  {\bf 26}, 131 (2023).

\bibitem{bubenik2020persistent}
P.~Bubenik, M.~Hull, D.~Patel, B.~Whittle, {\it Inverse Problems\/} {\bf 36},
  025008 (2020).

\bibitem{craddock2012whole}
R.~C. Craddock, G.~A. James, P.~E. Holtzheimer~III, X.~P. Hu, H.~S. Mayberg,
  {\it Human brain mapping\/} {\bf 33}, 1914 (2012).

\bibitem{jiang2021fundamental}
X.~Jiang, T.~Zhang, S.~Zhang, K.~M. Kendrick, T.~Liu, {\it Psychoradiology\/}
  {\bf 1}, 23 (2021).

\bibitem{henselman1606matroid}
G.~Henselman, R.~Ghrist, {\it arXiv preprint arXiv:1606.00199\/} .

\bibitem{ghrist2014elementary}
R.~W. Ghrist, {\it Elementary applied topology\/}, vol.~1 (Createspace Seattle,
  2014).

\bibitem{ghrist2008barcodes}
R.~Ghrist, {\it Bulletin of the American Mathematical Society\/} {\bf 45}, 61
  (2008).

\bibitem{harris2020array}
C.~R. Harris, {\it et~al.\/}, {\it Nature\/} {\bf 585}, 357 (2020).

\bibitem{sedgwick2012multiple}
P.~Sedgwick, {\it Bmj\/} {\bf 344} (2012).

\bibitem{fisher1992statistical}
R.~A. Fisher, {\it Breakthroughs in statistics\/} (Springer, 1992), pp. 66--70.

\bibitem{frazier2018tutorial}
P.~I. Frazier, {\it arXiv preprint arXiv:1807.02811\/}  (2018).

\bibitem{williams2006gaussian}
C.~K. Williams, C.~E. Rasmussen, {\it Gaussian processes for machine
  learning\/}, vol.~2 (MIT press Cambridge, MA, 2006).

\bibitem{balandat2020botorch}
M.~Balandat, {\it et~al.\/}, {\it Advances in neural information processing
  systems\/} {\bf 33}, 21524 (2020).

\bibitem{praturu2024adaptive}
A.~Praturu, T.~O. Sharpee, {\it Iscience\/} {\bf 27} (2024).

\bibitem{schwieterman2014strawberry}
M.~L. Schwieterman, {\it et~al.\/}, {\it PloS one\/} {\bf 9}, e88446 (2014).

\bibitem{wang2015theta}
Y.~Wang, S.~Romani, B.~Lustig, A.~Leonardo, E.~Pastalkova, {\it Nature
  neuroscience\/} {\bf 18}, 282 (2015).

\bibitem{pastalkova2008internally}
E.~Pastalkova, V.~Itskov, A.~Amarasingham, G.~Buzs{\'a}ki, {\it Science\/} {\bf
  321}, 1322 (2008).

\bibitem{glasser2013minimal}
M.~F. Glasser, {\it et~al.\/}, {\it Neuroimage\/} {\bf 80}, 105 (2013).

\end{thebibliography}

\bibliographystyle{Science}

\section*{Acknowledgments}
The authors thank Tatyana Sharpee for sharing code and helping to replicate the sampling of points from candidate hyperbolic spaces and Salar Fattahi for helpful discussions around Bayesian optimization.   Conceptualization: A.J.S.; Investigation: A.J.S.; Methodology: A.J.S., R.A.K.; Software: A.J.S., N.S., R.A.K.; Writing – original draft: A.J.S.; Writing – review \& editing: A.J.S., N.S., R.A.K., C.G., M.G.B.

\newpage
\section*{Online Methods}
\subsection*{Filtered simplicial complexes \& persistent homology}

To compute the elements of a filtered simplicial complex, TDA methods follow the same steps as classic network analyses, namely, thresholding and binarizing, but do so for all possible thresholds. Thresholding involves setting entries of a dissimilarity matrix, $a_{ij}$, that are below some threshold $t$ equal to $0$. Binarizing then sets the remaining non-zero entries equal to $1$. Importantly, and as mentioned above, in TDA this is done for all possible thresholds thereby retaining all of the information in the relative values of $A_{ij}$, while in network science, a small set of thresholds are typically chosen and analyses are conducted on the binarized matrices at each threshold. 

Each of these binarized matrices defines a simplicial complex, which is a set composed of the 0-simplices (the vertices of the graph), the 1-simplices (the edges of the graph), the 2-simplices (the faces formed by a cycle between three vertices), and so on. The filtered simplicial complex is the ordered set of simplicial complexes defined by the thresholds $t_i<t_{i+1}$ so that the simplicial complexes are nested subsets, $S_i\subset S_{i+1}$. Persistent homology computations are performed on the filtered simplicial complex to track the birth and death of topological features in the data as the threshold increases from $t_0$. The collection of these birth and death records across all thresholds for a specific dimension $k$ is referred to as a barcode and the Betti number at some threshold $t_i$ is the number of ``alive" topological features at $t_i$. See~\cite{henselman1606matroid,ghrist2014elementary,ghrist2008barcodes} for a detailed treatment.

\subsection*{Statistics on Betti curves}
Betti curves were evaluated against each other by two statistics~\cite{giusti_clique_2015,zhou2018hyperbolic}: the itegrated betti value or the area under the curve (ibv) and by l1-distances between curves. For a given Betti curve $\beta(t)$ the ibv is given by:\begin{equation}
    ibv = \int_0^1 \beta(t)dt
    \label{eq:ibv}
\end{equation} Similarly, for two different Betti curves, $\beta^a(t)$ and $\beta^b(t)$, the l1 distance between the two curves is given by the integral of the absolute difference between the two curves:\begin{equation}
    d_{l1} = \int_0^1 |\beta^a(t)-\beta^b(t)|dt
    \label{eq:l1d}
\end{equation} ALBATROSS uses numpy's~\cite{harris2020array} \textit{trapz} function to compute these integrals.

\subsection*{Statistical Inference}
We take advantage of the central limit theorem (CLT) to statistically infer the geometry of experimental data. Specifically, the mean and variance of measures on Betti curves, such as L1-distance and area under the curve~\cite{zhou2018hyperbolic,giusti_clique_2015} are estimated for best fit and shuffled geometric spaces and then used to test whether the empirical data are consistent with that space. Concretely, for a measure, $\mu$, on a Betti curve, the distribution of $\mu(\beta)$ is estimated as a normal distribution via the CLT: \begin{equation}
    \mu(\beta) \sim N(mean[\mu(\beta^r)], var[\mu(\beta^r)]*r)=N(\bar{\mu},\sigma^2_\mu)
    \label{eq:norm}
\end{equation} from $r$ re-samplings from the $I$ iterations.

Once the mean and variance of a given measure on Betti curves are estimated by re-sampling of generated Betti curves (\ref{eq:norm}), a test statistic is computed by comparing the mean empirical betti curve measure, $\overline{\mu_{emp}}$ to the distribution estimated from a candidate geometric model or a shuffled adjacency matrix:
\begin{equation}
    Z_\mu = \frac{\overline{\mu_{emp}}-\bar{\mu}}{\sqrt{\sigma^2_\mu}}
    \label{eq:z}
\end{equation} These $Z$ statistics are computed for the L1-distance between empirical and best fit geometric Betti curves \textit{and} for the area under the Betti curves~\cite{zhou2018hyperbolic}. This results in six $Z$ statistics for each candidate space. These $Z$ statistics are then converted into two-tailed p-values and corrected for six multiple comparisons with the Bonferroni method~\cite{sedgwick2012multiple}.

Finally, the corrected p-values are used to generate a chi squared statistic via Fisher's method~\cite{fisher1992statistical}:
\begin{equation}
    \chi^2_{2*6} \sim -2\sum_{j=1}^{6} log(p_j)
\end{equation}

\subsection*{Euclidean Geometric Model}
Euclidean models were specified by their number of dimensions, $d$, and by a noise level, $\epsilon$~\cite{zhou2018hyperbolic}. For a given sub-sample size, $n$, the same number of points, $n$, were randomly sampled from the $d$-dimensional unit cube. Distances between the $n$ points were calculated via the standard euclidean distance:\begin{equation}
    d_{ij} = \sqrt{\sum_{k=1}^{d} (x^i_k-x^j_k)^2}
\end{equation} Finally, multiplicative noise distributed according to $N(0,\epsilon)$ was symmetrically added to the distance matrix.

\subsection*{Hyperbolic Geometric Model}
Hyperbolic models were specified by their number of dimensions, $d$, by a noise level, $\epsilon$, by a maximum radius, $r_{max}$, and by a minimum radius percentage, $r_{min} \in [0,1]$~\cite{zhou2018hyperbolic}. For a given sub-sample size, $n$, the same number of points, $n$, were randomly sampled from the $d$-dimensional hyperbolic spherical shell with maximum radius $r_{max}$, minimum radius $r_{min}*r_{max}$. This sampling started with sampling $n$ points within the euclidean $d$ dimensional unit cube and sampling $n$ radii $\in [r_{max}*r_{min}, r_{max}]$ via the distribution:\begin{equation}
    p(r) \sim sinh((d-1)*r)
\end{equation} This sampling distribution ensures random uniform sampling of radii in hyperbolic models where space expands away from the origin~\cite{zhou2018hyperbolic} (i.e. there is negative curvature).

Next the pairwise angles, $\theta_{ij}$, between the $n$ points sampled from $d$-dimensional euclidean space are computed via the dot-product: \begin{equation}
    cos(\theta_{ij}) = \frac{\mathbf{x_i} \cdot \mathbf{x_j}}{||\mathbf{x_i}||||\mathbf{x_i}||}
\end{equation} where $||\mathbf{x}||$ is the l2 norm of $\mathbf{x}$. Next, the pairwise distance between sampled points in hyperbolic space is computed as:\begin{equation}
    d_{ij} = acosh(cosh(r_i)*cosh(r_j)-sinh(r_i)*sinh(r_j)*cos(\theta_{ij}))
    \label{eq:dist}
\end{equation} Finally, multiplicative noise distributed according to $N(0,\epsilon)$ was symmetrically added to the distance matrix.

\subsection*{Grid search for best fit euclidean geometry} We implemented a grid search in order to determine the best fit euclidean space to empirical Betti curves. For each point in the grid, $I$ Betti curves were generated from $n$ points sampled from the candidate space and averaged. In order to evaluate the fit of the candidate spaces at each grid point we first calculated the difference between the empirical and model integrated Betti values according to Equation \ref{eq:ibv}. We next calculated the l1-distance between the average empirical Betti curves and average model curves according to Equations \ref{eq:l1d}. Since the first three Betti curves often have different shapes, locations, and areas under them depending on the source of the empirical data or the model parameters, we normalized these measures so that the fit to each Betti curve has close to equal importance in selecting a best fit model. Specifically, for the ibv distances we normalized by the ibv of the empirical Betti curves. Similarly, we normalized the l1-distances by the l1-distance between each empirical Betti curve and its distance from the horizontal line defined by the mean Betti value:
\begin{equation}
    l1^k = \frac{\int_0^1 |\beta^k_{emp}(t)-\beta^k_{model}(t)|dt}{\int_0^1 |\beta^k_{emp}(t)-\overline{\beta^k_{emp}(t)}|dt}
\end{equation}
\begin{equation}
    ibv^k = \frac{|\int_0^1 \beta^k_{emp}(t)dt - \int_0^1 \beta^k_{model}(t)dt|}{\int_0^1 \beta^k_{emp}(t)dt}
\end{equation} Finally, the grid point with the smallest combined metric, $\sum_{k=1}^3 l1^k + ibv^k$, is chosen as the best fit model. 

The default grid for the euclidean model was $d\in[1,100]$, and $\epsilon \in[0.0,0.15]$. When best fit models were chosen with parameter values on the extremes of the grid ranges, the protocol was re-run with a larger grid to ensure that a large enough space was explored to select the best fitting geometric model. 

\subsection*{Bayesian optimization for best fit hyperbolic geometry}

Unlike Euclidean models, where the ($d$, $\epsilon$) search space is only two dimensional, selecting the best-fit hyperbolic geometry is more challenging. There are three hyper-parameters in the hyperbolic model ($\epsilon$, $r_{max}$, $r_{min}$). Since Betti curves depend on these hyperparameters in a highly nonlinear manner, directly estimating the optimal configuration is difficult. Moreover, as computing a single Betti curve requires up to 4000 seconds, it is essential to minimize the number of function evaluations. To address this challenge, we employed Bayesian optimization (BO) to efficiently explore the hyperbolic parameter space and identify configurations that best replicate the empirical Betti curves.

Our objective function is defined as:

\begin{equation}
    \text{argmax}_{(\epsilon, r_{min},r_{max})} \text{obj} = -\sum_{k=1}^3 l_1^k+ibv^k
\end{equation}

,
where $l_1^k$  and $ibv^k$ are defined in Equations 10 and 11. The construction of Betti curves relies on simplex counting, which is inherently data-dependent and non-differentiable. Consequently, the objective function exhibits a highly nonlinear dependence on the hyperparameters ($\epsilon$, $r_{max}$, $r_{min}$) and is non-differentiable, making direct optimization challenging.

To address this, BO provides a model-based, derivative-free optimization approach designed for expensive black-box functions~\cite{frazier2018tutorial}. We use a Gaussian process (GP) surrogate model to approximate the mapping from hyperparameters ($\epsilon$, $r_{max}$, $r_{min}$, obj) to the objective value. GP surrogates provide a flexible nonparametric estimate of the objective landscape from hyper-parameters to objectives. In our experiments, we instantiate the GP using a Matérn kernel, a standard choice for modeling non-smooth functions. Kernel parameters, including length-scale and variance, are optimized by maximizing the GP log-marginal likelihood at each iteration.

To select new candidates, we use the expected improvement (EI) acquisition function~\cite{williams2006gaussian}. As its name suggests, EI is the expected improvement over maximum objective value from existing observations. EI naturally balances exploration of uncertain regions with exploitation near promising areas. At each iteration, the acquisition function is maximized to propose the next parameter configuration ($\epsilon$, $r_{max}$, $r_{min}$, obj) to evaluate.

The BO procedure is implemented using the BoTorch package~\cite{balandat2020botorch} built on PyTorch, which supports scalable GP surrogates and flexible acquisition functions. We use the default optimizer settings in BoTorch for maximizing EI and limited the optimization to 100 iterations. For initialization, we employ Latin Hypercube Sampling (LHS) to generate five initial design points uniformly across the search space, defined as:

\begin{equation}
    \epsilon \in [10^{-6},1], r_{max} \in [10^{-6},1], r_{min} \in [10^{-6},.99]
\end{equation}

\subsection*{Distribution of topological properties across human cortex}
We ran ALBATROSS with $n$=35 and $I$=150 for each parcel and subject from the HCP data, using -1 times the absolute value of the correlation matrix as the input. We ran this for voxelwise resting state functional correlation matrices (rsFC) for 392 previously defined parcels and for the 392x392 matrix of mean timecourses from each parcel. The parameters and statistics for the voxelwise fits for the 392 parcels are in table (Supplementary Table \ref{tab:fitsCC}).

\subsection*{Hyperbolic Multi Dimensional Scaling}
In order to determine coordinates for the 392 parcels in 3-D hyperbolic space, we ran Bayesian hyperbolic multi dimensional scaling (hMDS) in the n-dimensional Poincare Ball. The code for hMDS is available on github (\url{https://github.com/enlberman/albatross}) and was adapted from existing code (\url{https://github.com/sharpee/BayesianHMDS}) for bayesian hMDS based on lorentz coordinates~\cite{praturu2024adaptive}.

\subsection*{Experimental Data}
\subsubsection*{Strawberry odor data}
Concentrations of individual monomolecular odors from different cultivars of strawberries~\cite{schwieterman2014strawberry} were obtained from: https://doi.org/10.1371/journal.pone.0088446.s005. A matrix representing the distance between different monomolecular odors across strawberry samples was computed as -1 times the absolute value of correlation matrix so that monomolecular odors that were more similar had lower values:
\begin{equation}
c_{ij} = -\left|\frac{\sum(x_i-\bar{x})(y_i-\bar{y})}{\sqrt{\sum(x_i-\bar{x})^2}\sqrt{\sum(y_i-\bar{y})^2}}\right|
\end{equation}

\subsubsection*{Hippocampal place cell data}
Spike trains of neurons in area CA1 of rodent hippocampus were recorded during spatial navigation in a familiar, 2D, 1.5 m x 1.5 m square box environment. Experimental procedures and computation of pairwise correlations have been previously described in detail in ~\cite{giusti_clique_2015,wang2015theta,pastalkova2008internally}. Briefly,  cross-correlograms were computed as $ccg_{ij}(\tau) = \frac{1}{T}\int_0^Tf_i(t)f_j(t+\tau)dt$ where $f_i(t)$ is the firing rate of the ith neuron and $T$ to total duration of the considered time period. Since larger values corresponded to more synchronized neurons, $-ccg_{ij}$ was used as the input empirical distance matrix to ALBATROSS.

\subsubsection*{Human Connectome Project fMRI data}
    We analyzed data from the Human Connectome Project (HCP), a multi-site consortium that collected MRI, behavioral, and demographic data from 1,022 subjects. We analyzed FIX-ICA denoised resting state data provided by the Human Connectome Project. The acquisition parameters and prepossessing of these data have been described in detail elsewhere \cite{glasser2013minimal}. Briefly, this preprocessing included distortion correction, realignment, transformation to a standard space, high-pass filtering, and ICA-FIX de-noising to remove motion artifacts.
    
    We randomly chose 10 subjects from those with both the left-to-right and right-to-left phase encoding scans available from the first resting state session. We additionally restricted the choice to subjects where the mean framewise displacement was less than .2mm and the maximum framewise displacement was less than 2mm for both scans. The two scans for each subject were averaged, and then voxel-wise correlation matrices were obtained within 392 previously defined functional parcels~\cite{craddock2012whole}. These parcels have a minimum of 56 voxels, a maximum of 600 voxels, a mean of 370 voxels, and a standard deviation of 78.8 voxels.

\newpage
\clearpage
\section*{Supplementary Figures}
\setcounter{figure}{0}
\renewcommand{\figurename}{Supplementary Figure}

\begin{figure}[hbpt!]
\centering
\includegraphics[width=1\textwidth]{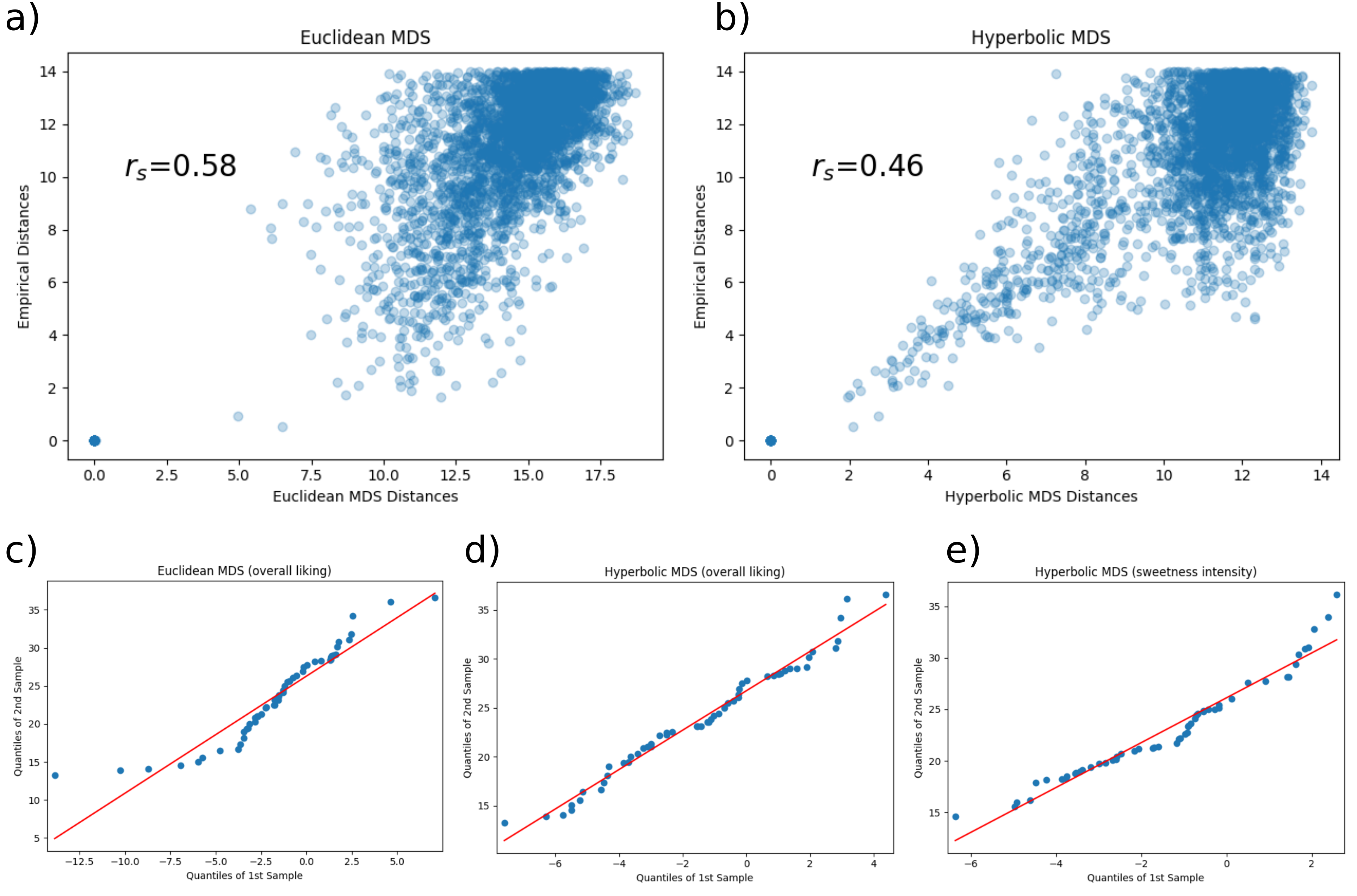}
\caption{Model Comparisons between Euclidean and hyperbolic multi dimensional scaling. a),b) Both Euclidean and Hyperbolic multi dimensional scaling found reasonable embeddings. c) quantile-quantile plots of the residuals from the discovered perceptual axis in euclidean space demonstrate the presence of remaining nonlinearity in the data and suggesting that a linear axis is not appropriate. d),e) In contrast, the linear axes found in Hyperbolic space are well fit suggesting that the transformation of data and distances in that space has resulted in a linear relationship between monomolecular odors and perceptual ratings.}
\label{fig:mdsCompStats}
\end{figure}
\newpage

\newpage
\begin{figure}[hbpt!]
\centering
\includegraphics[width=1\textwidth]{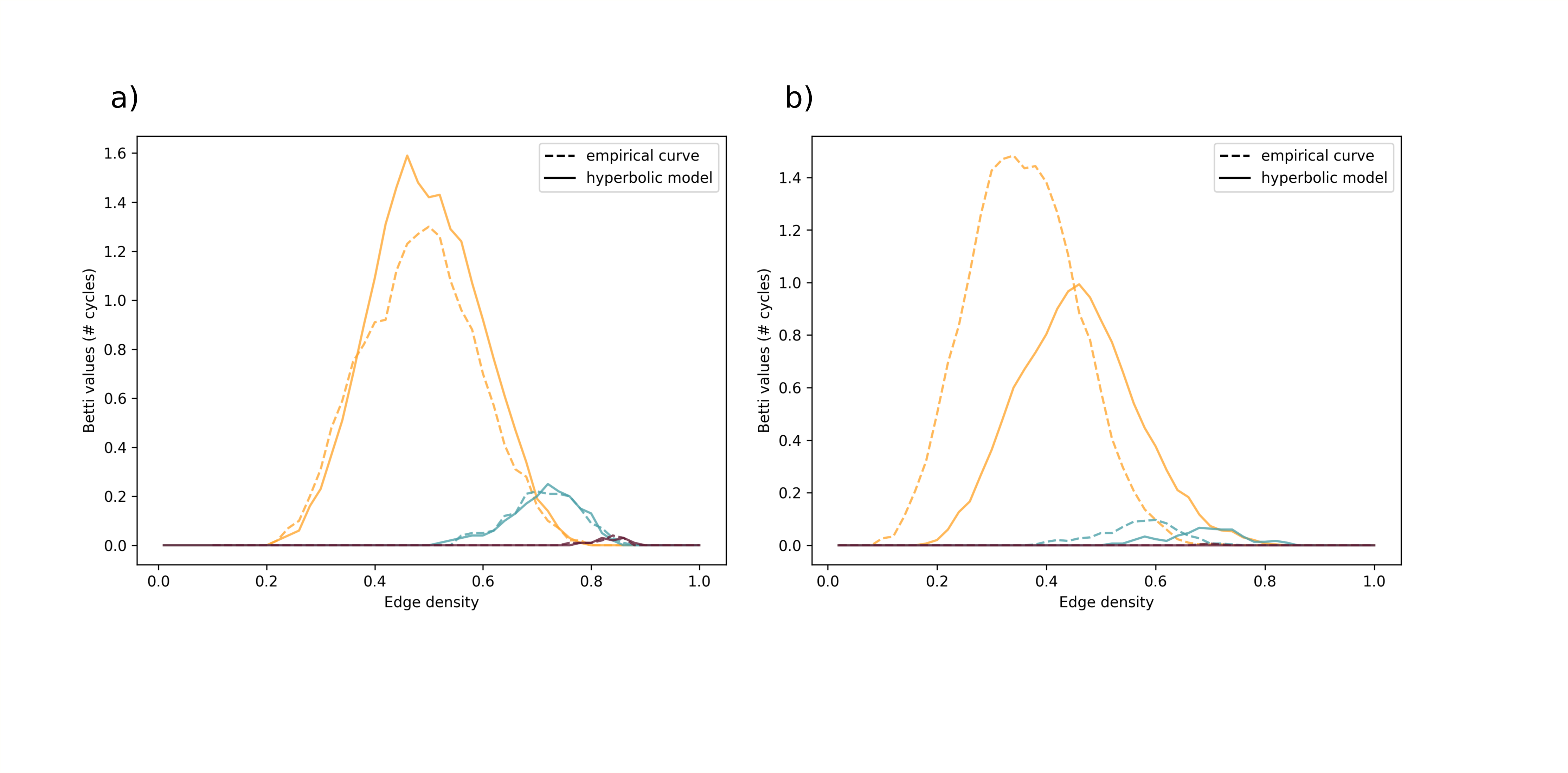}
\caption{ALBATROSS is able to find visually well fitting geometries even with sub-sample sizes of 10 though the small number of possible cycles in the filrations at mean that there is too much variance at these levels for robust statistical inference. a) Strawberry odors. b) Hippocampal place cells.}
\label{fig:smallmodels}
\end{figure}

\newpage
\begin{figure}[hbpt!]
\centering
\includegraphics[width=1\textwidth]{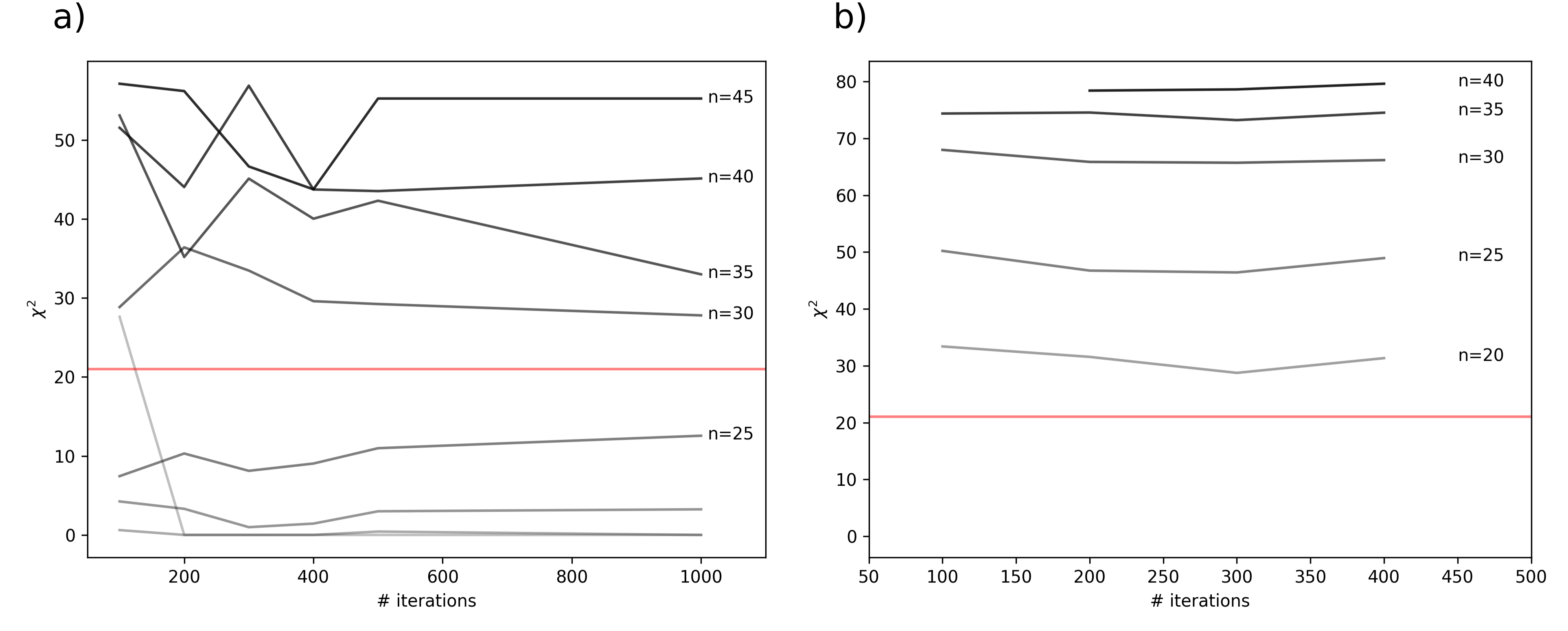}
\caption{Above and including a sub-sample size of 30, one hundred iterations are sufficient for statistical inference of global geometric structure in both the strawberry odor a) and hippocampal place neuron b) datasets.}
\label{fig:iters}
\end{figure}

\newpage
\begin{figure}[hbpt!]
\centering
\includegraphics[width=1\textwidth]{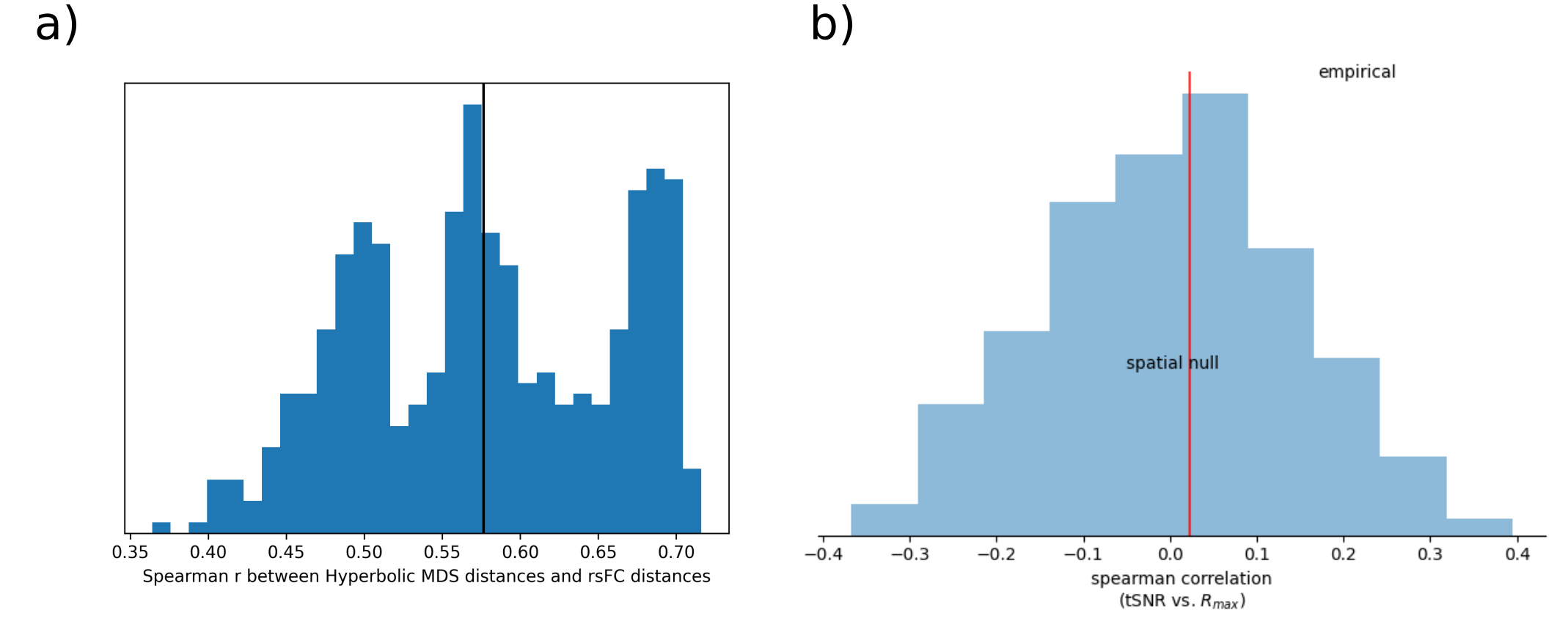}
\caption{HCP data sensitivity checks. a) The strong correlation between embedded and empirical distances for the 392x392 rsFC matrices indicates that the hMDS algorithm found good embedding. b) Compared to a spatial null model that preserves spatial autocorrelation, the pattern of hyperbolic radii over the human cortex is not significantly correlated with the temporal signal to noise ratio (tSNR) indicating that the patterns of topological properties across the cortex are not the result of spatially varying noise levels.}
\label{fig:noise}
\end{figure}

\newpage
\section*{Supplementary Tables}
\renewcommand{\tablename}{Supplementary Table}

\begin{table}[hbpt!]
\caption{Summary of median elapsed compute time in seconds, median memory usage in GB and the number of runs out of 16 that exceeded available memory on a 475 GB compute node for different sub-sample sizes, n.}
    \centering
  \begin{tabular}{rrrr}
  n &  Elapsed (s) &    MaxRSS (GB) &  \# Out of Memory \\
  \hline \hline
  50 &    301.0 &   0.552240 &                0 \\
100 &    305.0 &   1.083216 &                0 \\
200 &    381.0 &  15.104802 &                0 \\
225 &    416.0 &  24.985918 &                0 \\
250 &    487.5 &  31.338518 &                0 \\
300 &    733.0 &  70.517164 &                0 \\
325 &    866.5 &  97.117124 &                1 \\
400 &   1801.5 & 219.333704 &                2 \\
500 &   3942.0 & 470.140316 &               16 \\
550 &   4145.0 & 383.055260 &               16 \\
\hline \hline
\end{tabular}
    
    \label{tab:memory}
\end{table}

\begin{table}[hbpt!]
\caption{ALBATROSS Computation statistics on a Laptop Computer with 8-cores and 24 GB of RAM. The protocol was run on a random dataset of size 1,000.}
        \centering
  \begin{tabular}{rrrr}
  n &  iterations &   Elapsed (s) & MaxRSS (GB)\\
  \hline \hline
 30 & 100 &   5,953 &  1.5 \\
 \hline \hline
\end{tabular}
    \label{tab:pc}
\end{table}

\begin{table}[hbpt!]
\caption{Summary of different best fit hyperbolic model parameters for the strawberry odor data across different choices of $n\in[10,15,20,25,40,45,50,45,50,55]$ and $i\in[100,200,300,400,500,1000]$. Note that while different choices resulted in different model parameters and noise levels for the best fit model, all best fit models were hyperbolic shells, consistent with~\cite{zhou2018hyperbolic}. Best fit noise levels, set by $\epsilon$ varied considerably with $i$ and $n$. }
    \centering
  \begin{tabular}{rrrr}
  $r_{max}$ &  $r_{min}$ & \# of models & \% of models\\
  \hline \hline
 2 &    0.9 &   1  &  1.7\%  \\       
 3 &    0.8 &   4  &  6.9\%  \\       
 3 &    0.9 &   2  &  3.4\%  \\       
 4 &    0.8 &   1  &  1.7\%  \\       
 4 &    0.9 &   4  & 6.9\%    \\       
 4 &    1.0 &   1  & 1.7\%    \\       
 \textbf{5} &    \textbf{0.9} &   \textbf{37}  & \textbf{63.8\%}   \\       
 6 &    0.9 &  8   &  13.8\%  \\    
 \hline \hline
\end{tabular}
    
    \label{tab:modelparmsStraw}
\end{table}

\begin{table}[hbpt!]
\caption{Summary of different best fit hyperbolic model parameters for the hippocampal place cell data across different choices of $n\in[20,25,30,35,40]$ and $i\in[100,200,300,400]$. Both hyperbolic and euclidean spaces were consistent with the data. Note that while different choices resulted in different model parameters and noise levels for the best fit model, all best fit hyperbolic models were close to full hyperbolic spaces (i.e., not shells). Best fit noise levels, set by $\epsilon$ varied considerably with $i$ and $n$. Noise levels for euclidean models was generally low. The most common best fit model of each type is highlighted in bold.}
    \centering
  \begin{tabular}{rrrr}
  \hline \hline
  hyperbolic & & &\\
  \hline 
  $r_{max}$ &  $r_{min}$ & \# of models & \% of models\\
  \hline 
  
 1 &    0.0 &   8 &  33\% \\       
1 &    0.10 &   4  &  26\% \\

2 &    0.0 &   4  &  20\% \\  
 2 &    0.10 &   2  &  13\% \\       
 3 &    0.10 &   3  &  13\% \\  
 3 &    0.20 &   2  &  7\% \\  
 6 &    0.00 &   1  &   7\%  \\       
 \hline \\
 euclidean & & &\\
 \hline
 $\epsilon$ & dimension range  & \# of models & \% of models\\
  \hline 
 0.0 & 37-64    &   8  & 40\%  \\       
 0.01 & 22-52    &   8  & 40\%   \\       
 0.02 & 28    &   1  & 7\%  \\       
 0.03 & 28    &   1  &  7\% \\ 
 0.06 & 16    &   1  &  7\% \\ 
 0.09 &  12   &   1  &    7\% \\ 
 0.05 & 19    &   1  &   7\%  \\ 
 \hline \hline

\end{tabular}
    
    \label{tab:modelparmsHippo}
\end{table}

\newpage
\begin{longtable}{r|r|r|r|r|r|r}
\caption{Group-level fit statistics for voxelwise rsFC matrices for each of the 392 previously defined brain parcels.}\\
\label{tab:fitsCC} 
 label &  Hbp p-value &  Hbp $\chi^2$ &  Euc p-value &  Euc $\chi^2$  &  Shuffled p-value &  Shuffled $\chi^2$ \\
\hline
     1.0 & 358.536 & 0.0 & 333.997 & 0.0 & 197.453 & 0.0 \\
2.0 & 358.747 & 0.0 & 401.832 & 0.0 & 532.522 & 0.0 \\
3.0 & 292.041 & 0.0 & 364.117 & 0.0 & 422.298 & 0.0 \\
4.0 & 280.597 & 0.0 & 225.109 & 0.0 & 173.352 & 0.001 \\
5.0 & 462.738 & 0.0 & 517.0 & 0.0 & 462.019 & 0.0 \\
6.0 & 388.049 & 0.0 & 530.526 & 0.0 & 515.278 & 0.0 \\
7.0 & 257.745 & 0.0 & 287.489 & 0.0 & 396.311 & 0.0 \\
8.0 & 453.796 & 0.0 & 584.286 & 0.0 & 595.318 & 0.0 \\
9.0 & 413.609 & 0.0 & 254.06 & 0.0 & 178.446 & 0.0 \\
10.0 & 452.564 & 0.0 & 213.713 & 0.0 & 243.331 & 0.0 \\
11.0 & 331.781 & 0.0 & 397.895 & 0.0 & 334.687 & 0.0 \\
12.0 & 312.646 & 0.0 & 189.834 & 0.0 & 151.204 & 0.028 \\
13.0 & 240.971 & 0.0 & 327.786 & 0.0 & 351.068 & 0.0 \\
14.0 & 223.152 & 0.0 & 289.18 & 0.0 & 220.078 & 0.0 \\
15.0 & 354.498 & 0.0 & 529.14 & 0.0 & 445.332 & 0.0 \\
16.0 & 202.855 & 0.0 & 337.837 & 0.0 & 418.463 & 0.0 \\
17.0 & 269.525 & 0.0 & 373.366 & 0.0 & 233.211 & 0.0 \\
18.0 & 497.644 & 0.0 & 273.164 & 0.0 & 187.917 & 0.0 \\
19.0 & 387.698 & 0.0 & 353.756 & 0.0 & 179.879 & 0.0 \\
20.0 & 351.577 & 0.0 & 431.311 & 0.0 & 247.931 & 0.0 \\
21.0 & 337.608 & 0.0 & 675.433 & 0.0 & 183.314 & 0.0 \\
22.0 & 181.326 & 0.0 & 185.766 & 0.0 & 318.26 & 0.0 \\
23.0 & 250.953 & 0.0 & 467.791 & 0.0 & 524.345 & 0.0 \\
24.0 & 337.891 & 0.0 & 484.176 & 0.0 & 480.974 & 0.0 \\
25.0 & 248.679 & 0.0 & 463.352 & 0.0 & 149.884 & 0.034 \\
26.0 & 502.889 & 0.0 & 518.243 & 0.0 & 588.183 & 0.0 \\
27.0 & 190.109 & 0.0 & 493.163 & 0.0 & 172.725 & 0.001 \\
28.0 & 117.143 & 0.557 & 262.582 & 0.0 & 235.476 & 0.0 \\
29.0 & 305.996 & 0.0 & 321.803 & 0.0 & 205.698 & 0.0 \\
30.0 & 451.441 & 0.0 & 311.491 & 0.0 & 387.062 & 0.0 \\
31.0 & 221.496 & 0.0 & 284.617 & 0.0 & 379.249 & 0.0 \\
32.0 & 382.256 & 0.0 & 404.058 & 0.0 & 299.814 & 0.0 \\
34.0 & 178.357 & 0.0 & 339.311 & 0.0 & 281.449 & 0.0 \\
35.0 & 163.533 & 0.005 & 340.168 & 0.0 & 360.331 & 0.0 \\
36.0 & 336.109 & 0.0 & 355.032 & 0.0 & 295.171 & 0.0 \\
37.0 & 323.852 & 0.0 & 468.461 & 0.0 & 492.763 & 0.0 \\
38.0 & 406.006 & 0.0 & 237.375 & 0.0 & 159.055 & 0.01 \\
39.0 & 466.722 & 0.0 & 353.161 & 0.0 & 150.057 & 0.033 \\
40.0 & 199.68 & 0.0 & 231.329 & 0.0 & 307.848 & 0.0 \\
41.0 & 18.909 & 1.0 & 36.4 & 1.0 & 173.435 & 0.001 \\
42.0 & 17.353 & 1.0 & 44.335 & 1.0 & 163.357 & 0.005 \\
43.0 & 80.966 & 0.998 & 448.963 & 0.0 & 538.271 & 0.0 \\
44.0 & 32.015 & 1.0 & 189.456 & 0.0 & 468.197 & 0.0 \\
45.0 & 43.56 & 1.0 & 88.136 & 0.987 & 193.097 & 0.0 \\
46.0 & 26.707 & 1.0 & 126.08 & 0.334 & 282.3 & 0.0 \\
47.0 & 24.369 & 1.0 & 91.786 & 0.974 & 246.464 & 0.0 \\
48.0 & 8.73 & 1.0 & 102.654 & 0.872 & 289.169 & 0.0 \\
49.0 & 33.205 & 1.0 & 284.119 & 0.0 & 470.446 & 0.0 \\
50.0 & 23.296 & 1.0 & 57.388 & 1.0 & 172.508 & 0.001 \\
51.0 & 79.665 & 0.998 & 455.537 & 0.0 & 584.478 & 0.0 \\
52.0 & 8.312 & 1.0 & 167.085 & 0.003 & 334.232 & 0.0 \\
53.0 & 36.827 & 1.0 & 299.66 & 0.0 & 449.996 & 0.0 \\
54.0 & 14.443 & 1.0 & 192.288 & 0.0 & 405.658 & 0.0 \\
55.0 & 18.623 & 1.0 & 50.588 & 1.0 & 190.221 & 0.0 \\
56.0 & 34.128 & 1.0 & 218.517 & 0.0 & 388.689 & 0.0 \\
57.0 & 21.239 & 1.0 & 254.323 & 0.0 & 416.391 & 0.0 \\
58.0 & 19.387 & 1.0 & 208.929 & 0.0 & 411.89 & 0.0 \\
59.0 & 5.243 & 1.0 & 172.543 & 0.001 & 333.183 & 0.0 \\
60.0 & 3.808 & 1.0 & 72.927 & 1.0 & 214.931 & 0.0 \\
61.0 & 23.812 & 1.0 & 246.201 & 0.0 & 430.45 & 0.0 \\
62.0 & 110.832 & 0.714 & 332.668 & 0.0 & 491.099 & 0.0 \\
63.0 & 71.967 & 1.0 & 273.842 & 0.0 & 561.506 & 0.0 \\
64.0 & 8.489 & 1.0 & 199.249 & 0.0 & 391.754 & 0.0 \\
65.0 & 40.463 & 1.0 & 333.705 & 0.0 & 482.562 & 0.0 \\
66.0 & 11.673 & 1.0 & 223.15 & 0.0 & 440.189 & 0.0 \\
67.0 & 21.346 & 1.0 & 66.318 & 1.0 & 198.312 & 0.0 \\
68.0 & 104.269 & 0.846 & 409.473 & 0.0 & 543.171 & 0.0 \\
69.0 & 5.996 & 1.0 & 182.622 & 0.0 & 452.047 & 0.0 \\
70.0 & 14.808 & 1.0 & 161.899 & 0.007 & 357.849 & 0.0 \\
71.0 & 15.688 & 1.0 & 46.117 & 1.0 & 187.924 & 0.0 \\
72.0 & 20.258 & 1.0 & 125.717 & 0.342 & 290.638 & 0.0 \\
73.0 & 21.57 & 1.0 & 304.563 & 0.0 & 465.488 & 0.0 \\
74.0 & 16.844 & 1.0 & 170.013 & 0.002 & 328.458 & 0.0 \\
75.0 & 16.253 & 1.0 & 123.731 & 0.389 & 340.101 & 0.0 \\
76.0 & 7.095 & 1.0 & 121.143 & 0.454 & 266.603 & 0.0 \\
77.0 & 17.91 & 1.0 & 100.388 & 0.903 & 227.375 & 0.0 \\
78.0 & 12.095 & 1.0 & 39.063 & 1.0 & 156.945 & 0.013 \\
80.0 & 9.544 & 1.0 & 204.02 & 0.0 & 490.22 & 0.0 \\
81.0 & 50.608 & 1.0 & 326.45 & 0.0 & 501.431 & 0.0 \\
82.0 & 16.701 & 1.0 & 43.733 & 1.0 & 199.224 & 0.0 \\
83.0 & 9.627 & 1.0 & 175.141 & 0.001 & 390.812 & 0.0 \\
84.0 & 26.281 & 1.0 & 233.196 & 0.0 & 449.423 & 0.0 \\
85.0 & 22.193 & 1.0 & 45.204 & 1.0 & 159.031 & 0.01 \\
87.0 & 73.925 & 1.0 & 179.939 & 0.0 & 354.295 & 0.0 \\
88.0 & 43.043 & 1.0 & 242.182 & 0.0 & 417.302 & 0.0 \\
89.0 & 5.108 & 1.0 & 184.861 & 0.0 & 381.112 & 0.0 \\
90.0 & 4.396 & 1.0 & 124.206 & 0.378 & 301.819 & 0.0 \\
91.0 & 1.541 & 1.0 & 17.275 & 1.0 & 149.713 & 0.034 \\
92.0 & 2.871 & 1.0 & 163.988 & 0.005 & 351.218 & 0.0 \\
93.0 & 21.193 & 1.0 & 238.306 & 0.0 & 436.872 & 0.0 \\
94.0 & 45.039 & 1.0 & 345.656 & 0.0 & 546.284 & 0.0 \\
95.0 & 70.585 & 1.0 & 387.386 & 0.0 & 507.573 & 0.0 \\
96.0 & 27.217 & 1.0 & 218.412 & 0.0 & 386.637 & 0.0 \\
97.0 & 42.027 & 1.0 & 111.961 & 0.687 & 216.372 & 0.0 \\
98.0 & 36.878 & 1.0 & 50.741 & 1.0 & 191.747 & 0.0 \\
99.0 & 7.595 & 1.0 & 113.74 & 0.644 & 294.587 & 0.0 \\
100.0 & 3.632 & 1.0 & 32.02 & 1.0 & 180.361 & 0.0 \\
101.0 & 13.91 & 1.0 & 70.797 & 1.0 & 208.451 & 0.0 \\
102.0 & 14.206 & 1.0 & 253.386 & 0.0 & 428.036 & 0.0 \\
103.0 & 27.243 & 1.0 & 90.875 & 0.978 & 230.6 & 0.0 \\
104.0 & 32.394 & 1.0 & 281.392 & 0.0 & 408.97 & 0.0 \\
105.0 & 42.71 & 1.0 & 356.366 & 0.0 & 534.214 & 0.0 \\
106.0 & 19.031 & 1.0 & 188.417 & 0.0 & 429.915 & 0.0 \\
107.0 & 11.118 & 1.0 & 57.232 & 1.0 & 182.042 & 0.0 \\
108.0 & 69.108 & 1.0 & 406.634 & 0.0 & 534.233 & 0.0 \\
109.0 & 14.769 & 1.0 & 65.289 & 1.0 & 200.291 & 0.0 \\
110.0 & 11.367 & 1.0 & 151.027 & 0.029 & 322.945 & 0.0 \\
111.0 & 38.669 & 1.0 & 107.101 & 0.794 & 216.936 & 0.0 \\
112.0 & 18.339 & 1.0 & 47.445 & 1.0 & 192.465 & 0.0 \\
113.0 & 17.321 & 1.0 & 84.922 & 0.994 & 255.107 & 0.0 \\
114.0 & 31.296 & 1.0 & 394.46 & 0.0 & 536.449 & 0.0 \\
115.0 & 93.956 & 0.962 & 469.395 & 0.0 & 583.257 & 0.0 \\
116.0 & 18.006 & 1.0 & 197.273 & 0.0 & 424.692 & 0.0 \\
117.0 & 15.329 & 1.0 & 191.539 & 0.0 & 353.919 & 0.0 \\
118.0 & 8.396 & 1.0 & 120.073 & 0.481 & 289.855 & 0.0 \\
119.0 & 13.32 & 1.0 & 95.068 & 0.955 & 306.411 & 0.0 \\
120.0 & 83.23 & 0.996 & 457.953 & 0.0 & 591.745 & 0.0 \\
121.0 & 10.978 & 1.0 & 97.75 & 0.932 & 308.254 & 0.0 \\
122.0 & 25.916 & 1.0 & 309.244 & 0.0 & 461.775 & 0.0 \\
123.0 & 11.543 & 1.0 & 127.746 & 0.297 & 282.014 & 0.0 \\
124.0 & 21.129 & 1.0 & 122.492 & 0.42 & 265.42 & 0.0 \\
125.0 & 8.113 & 1.0 & 39.541 & 1.0 & 155.207 & 0.017 \\
126.0 & 40.135 & 1.0 & 97.348 & 0.936 & 207.167 & 0.0 \\
127.0 & 5.638 & 1.0 & 51.539 & 1.0 & 233.336 & 0.0 \\
128.0 & 6.479 & 1.0 & 102.621 & 0.872 & 276.525 & 0.0 \\
129.0 & 13.611 & 1.0 & 179.481 & 0.0 & 356.734 & 0.0 \\
130.0 & 24.606 & 1.0 & 61.16 & 1.0 & 173.438 & 0.001 \\
131.0 & 3.709 & 1.0 & 54.956 & 1.0 & 216.633 & 0.0 \\
132.0 & 16.204 & 1.0 & 53.726 & 1.0 & 180.425 & 0.0 \\
133.0 & 14.697 & 1.0 & 175.706 & 0.001 & 364.927 & 0.0 \\
134.0 & 13.359 & 1.0 & 77.208 & 0.999 & 261.226 & 0.0 \\
135.0 & 112.471 & 0.675 & 535.248 & 0.0 & 610.865 & 0.0 \\
136.0 & 4.714 & 1.0 & 137.046 & 0.137 & 357.728 & 0.0 \\
137.0 & 8.589 & 1.0 & 123.111 & 0.404 & 272.351 & 0.0 \\
138.0 & 93.926 & 0.962 & 412.58 & 0.0 & 588.538 & 0.0 \\
139.0 & 11.793 & 1.0 & 42.269 & 1.0 & 175.548 & 0.001 \\
140.0 & 13.497 & 1.0 & 108.49 & 0.766 & 276.963 & 0.0 \\
141.0 & 50.895 & 1.0 & 424.346 & 0.0 & 604.809 & 0.0 \\
142.0 & 36.9 & 1.0 & 145.332 & 0.058 & 343.511 & 0.0 \\
143.0 & 44.606 & 1.0 & 93.949 & 0.962 & 242.335 & 0.0 \\
144.0 & 11.868 & 1.0 & 33.362 & 1.0 & 161.208 & 0.007 \\
145.0 & 7.557 & 1.0 & 112.86 & 0.665 & 234.123 & 0.0 \\
146.0 & 4.168 & 1.0 & 36.617 & 1.0 & 177.418 & 0.001 \\
147.0 & 37.087 & 1.0 & 87.808 & 0.988 & 214.276 & 0.0 \\
148.0 & 1.252 & 1.0 & 39.797 & 1.0 & 175.173 & 0.001 \\
149.0 & 8.233 & 1.0 & 58.622 & 1.0 & 259.845 & 0.0 \\
150.0 & 13.475 & 1.0 & 90.981 & 0.978 & 244.364 & 0.0 \\
151.0 & 42.8 & 1.0 & 368.013 & 0.0 & 553.248 & 0.0 \\
152.0 & 13.318 & 1.0 & 30.729 & 1.0 & 161.73 & 0.007 \\
153.0 & 30.202 & 1.0 & 153.06 & 0.022 & 339.413 & 0.0 \\
154.0 & 39.173 & 1.0 & 239.335 & 0.0 & 404.425 & 0.0 \\
155.0 & 9.47 & 1.0 & 306.715 & 0.0 & 538.39 & 0.0 \\
156.0 & 14.186 & 1.0 & 187.393 & 0.0 & 411.526 & 0.0 \\
157.0 & 11.828 & 1.0 & 163.136 & 0.005 & 337.722 & 0.0 \\
158.0 & 16.266 & 1.0 & 135.674 & 0.155 & 304.103 & 0.0 \\
159.0 & 4.288 & 1.0 & 228.634 & 0.0 & 363.954 & 0.0 \\
160.0 & 30.475 & 1.0 & 211.186 & 0.0 & 423.954 & 0.0 \\
161.0 & 15.748 & 1.0 & 57.606 & 1.0 & 218.664 & 0.0 \\
162.0 & 8.689 & 1.0 & 61.854 & 1.0 & 199.566 & 0.0 \\
163.0 & 48.822 & 1.0 & 269.183 & 0.0 & 416.057 & 0.0 \\
164.0 & 111.723 & 0.693 & 527.031 & 0.0 & 594.643 & 0.0 \\
165.0 & 25.601 & 1.0 & 236.268 & 0.0 & 373.898 & 0.0 \\
166.0 & 3.946 & 1.0 & 14.973 & 1.0 & 151.634 & 0.027 \\
167.0 & 16.36 & 1.0 & 196.335 & 0.0 & 354.323 & 0.0 \\
168.0 & 4.565 & 1.0 & 15.941 & 1.0 & 149.081 & 0.037 \\
169.0 & 40.477 & 1.0 & 340.295 & 0.0 & 518.676 & 0.0 \\
170.0 & 18.942 & 1.0 & 75.886 & 0.999 & 185.155 & 0.0 \\
171.0 & 20.395 & 1.0 & 154.726 & 0.018 & 327.523 & 0.0 \\
172.0 & 6.509 & 1.0 & 230.629 & 0.0 & 555.271 & 0.0 \\
173.0 & 8.915 & 1.0 & 251.795 & 0.0 & 455.08 & 0.0 \\
174.0 & 13.817 & 1.0 & 71.179 & 1.0 & 238.162 & 0.0 \\
175.0 & 9.196 & 1.0 & 11.305 & 1.0 & 153.103 & 0.022 \\
176.0 & 23.383 & 1.0 & 292.723 & 0.0 & 516.617 & 0.0 \\
177.0 & 14.825 & 1.0 & 204.803 & 0.0 & 369.021 & 0.0 \\
178.0 & 17.741 & 1.0 & 220.985 & 0.0 & 376.08 & 0.0 \\
179.0 & 9.118 & 1.0 & 58.21 & 1.0 & 187.306 & 0.0 \\
180.0 & 9.938 & 1.0 & 115.311 & 0.604 & 277.771 & 0.0 \\
181.0 & 46.676 & 1.0 & 326.966 & 0.0 & 552.035 & 0.0 \\
182.0 & 24.715 & 1.0 & 87.795 & 0.988 & 224.384 & 0.0 \\
183.0 & 90.985 & 0.978 & 421.884 & 0.0 & 576.888 & 0.0 \\
184.0 & 22.764 & 1.0 & 103.665 & 0.856 & 272.07 & 0.0 \\
185.0 & 20.2 & 1.0 & 94.896 & 0.956 & 236.766 & 0.0 \\
186.0 & 16.894 & 1.0 & 251.066 & 0.0 & 411.284 & 0.0 \\
187.0 & 36.352 & 1.0 & 239.243 & 0.0 & 379.156 & 0.0 \\
188.0 & 18.032 & 1.0 & 38.375 & 1.0 & 174.851 & 0.001 \\
189.0 & 99.777 & 0.91 & 437.738 & 0.0 & 624.396 & 0.0 \\
190.0 & 15.335 & 1.0 & 217.336 & 0.0 & 447.645 & 0.0 \\
191.0 & 25.619 & 1.0 & 228.276 & 0.0 & 387.155 & 0.0 \\
192.0 & 22.128 & 1.0 & 357.722 & 0.0 & 560.842 & 0.0 \\
193.0 & 48.521 & 1.0 & 372.067 & 0.0 & 522.7 & 0.0 \\
194.0 & 47.101 & 1.0 & 288.849 & 0.0 & 459.036 & 0.0 \\
195.0 & 10.175 & 1.0 & 29.574 & 1.0 & 152.968 & 0.023 \\
196.0 & 4.008 & 1.0 & 39.905 & 1.0 & 168.31 & 0.002 \\
197.0 & 10.391 & 1.0 & 147.945 & 0.042 & 334.515 & 0.0 \\
198.0 & 29.418 & 1.0 & 304.603 & 0.0 & 469.166 & 0.0 \\
199.0 & 20.647 & 1.0 & 333.92 & 0.0 & 511.524 & 0.0 \\
200.0 & 5.803 & 1.0 & 92.205 & 0.972 & 250.528 & 0.0 \\
201.0 & 29.238 & 1.0 & 139.041 & 0.113 & 327.487 & 0.0 \\
202.0 & 10.696 & 1.0 & 147.225 & 0.046 & 366.019 & 0.0 \\
203.0 & 35.282 & 1.0 & 84.85 & 0.994 & 209.634 & 0.0 \\
204.0 & 8.916 & 1.0 & 109.631 & 0.741 & 338.904 & 0.0 \\
205.0 & 11.255 & 1.0 & 141.308 & 0.089 & 291.821 & 0.0 \\
206.0 & 91.761 & 0.974 & 456.388 & 0.0 & 575.669 & 0.0 \\
207.0 & 99.008 & 0.919 & 448.504 & 0.0 & 622.897 & 0.0 \\
208.0 & 12.425 & 1.0 & 47.862 & 1.0 & 177.72 & 0.0 \\
209.0 & 25.014 & 1.0 & 238.877 & 0.0 & 475.93 & 0.0 \\
210.0 & 13.14 & 1.0 & 189.134 & 0.0 & 345.718 & 0.0 \\
211.0 & 159.23 & 0.01 & 436.201 & 0.0 & 576.919 & 0.0 \\
212.0 & 21.953 & 1.0 & 274.695 & 0.0 & 442.908 & 0.0 \\
213.0 & 11.881 & 1.0 & 41.969 & 1.0 & 164.396 & 0.004 \\
214.0 & 18.977 & 1.0 & 221.022 & 0.0 & 372.032 & 0.0 \\
215.0 & 16.519 & 1.0 & 154.308 & 0.019 & 307.592 & 0.0 \\
216.0 & 18.873 & 1.0 & 173.243 & 0.001 & 358.013 & 0.0 \\
217.0 & 18.574 & 1.0 & 351.071 & 0.0 & 568.865 & 0.0 \\
218.0 & 38.819 & 1.0 & 358.608 & 0.0 & 520.995 & 0.0 \\
219.0 & 94.54 & 0.958 & 400.163 & 0.0 & 545.442 & 0.0 \\
220.0 & 25.909 & 1.0 & 165.965 & 0.003 & 322.053 & 0.0 \\
221.0 & 9.903 & 1.0 & 74.013 & 1.0 & 208.963 & 0.0 \\
222.0 & 15.482 & 1.0 & 187.409 & 0.0 & 390.84 & 0.0 \\
223.0 & 10.771 & 1.0 & 82.81 & 0.996 & 227.938 & 0.0 \\
224.0 & 2.984 & 1.0 & 40.728 & 1.0 & 194.229 & 0.0 \\
225.0 & 25.207 & 1.0 & 171.698 & 0.001 & 403.189 & 0.0 \\
226.0 & 26.188 & 1.0 & 139.001 & 0.113 & 325.327 & 0.0 \\
227.0 & 3.753 & 1.0 & 26.277 & 1.0 & 156.097 & 0.015 \\
228.0 & 5.211 & 1.0 & 136.737 & 0.141 & 319.537 & 0.0 \\
229.0 & 62.712 & 1.0 & 381.528 & 0.0 & 513.647 & 0.0 \\
230.0 & 130.461 & 0.242 & 287.182 & 0.0 & 514.816 & 0.0 \\
231.0 & 17.195 & 1.0 & 146.737 & 0.049 & 302.561 & 0.0 \\
232.0 & 27.644 & 1.0 & 133.218 & 0.193 & 327.301 & 0.0 \\
233.0 & 26.781 & 1.0 & 70.243 & 1.0 & 193.572 & 0.0 \\
234.0 & 37.815 & 1.0 & 305.376 & 0.0 & 455.938 & 0.0 \\
235.0 & 8.71 & 1.0 & 241.171 & 0.0 & 449.165 & 0.0 \\
236.0 & 45.063 & 1.0 & 318.351 & 0.0 & 458.462 & 0.0 \\
237.0 & 64.792 & 1.0 & 464.214 & 0.0 & 624.416 & 0.0 \\
238.0 & 5.87 & 1.0 & 20.072 & 1.0 & 152.632 & 0.024 \\
239.0 & 37.918 & 1.0 & 121.168 & 0.453 & 291.902 & 0.0 \\
240.0 & 25.785 & 1.0 & 332.777 & 0.0 & 588.075 & 0.0 \\
241.0 & 68.229 & 1.0 & 230.918 & 0.0 & 413.634 & 0.0 \\
242.0 & 11.384 & 1.0 & 48.528 & 1.0 & 183.43 & 0.0 \\
243.0 & 13.827 & 1.0 & 119.162 & 0.504 & 253.963 & 0.0 \\
244.0 & 40.241 & 1.0 & 155.692 & 0.016 & 321.637 & 0.0 \\
245.0 & 8.34 & 1.0 & 28.612 & 1.0 & 165.1 & 0.004 \\
246.0 & 7.391 & 1.0 & 42.404 & 1.0 & 184.932 & 0.0 \\
247.0 & 25.741 & 1.0 & 92.484 & 0.971 & 210.568 & 0.0 \\
248.0 & 24.588 & 1.0 & 181.612 & 0.0 & 354.844 & 0.0 \\
249.0 & 17.85 & 1.0 & 244.358 & 0.0 & 471.216 & 0.0 \\
250.0 & 53.824 & 1.0 & 315.207 & 0.0 & 462.347 & 0.0 \\
251.0 & 11.936 & 1.0 & 246.967 & 0.0 & 449.649 & 0.0 \\
252.0 & 145.368 & 0.057 & 449.897 & 0.0 & 627.774 & 0.0 \\
253.0 & 3.895 & 1.0 & 29.349 & 1.0 & 174.602 & 0.001 \\
254.0 & 12.049 & 1.0 & 158.922 & 0.01 & 367.433 & 0.0 \\
255.0 & 31.679 & 1.0 & 158.245 & 0.011 & 326.645 & 0.0 \\
256.0 & 21.446 & 1.0 & 149.407 & 0.036 & 337.672 & 0.0 \\
257.0 & 24.931 & 1.0 & 159.886 & 0.009 & 324.835 & 0.0 \\
258.0 & 2.039 & 1.0 & 19.331 & 1.0 & 155.724 & 0.016 \\
259.0 & 27.997 & 1.0 & 295.477 & 0.0 & 471.589 & 0.0 \\
260.0 & 0.0 & 1.0 & 3.648 & 1.0 & 152.408 & 0.024 \\
261.0 & 29.864 & 1.0 & 270.891 & 0.0 & 448.983 & 0.0 \\
262.0 & 7.232 & 1.0 & 37.788 & 1.0 & 167.083 & 0.003 \\
263.0 & 11.788 & 1.0 & 246.128 & 0.0 & 419.095 & 0.0 \\
264.0 & 33.03 & 1.0 & 179.632 & 0.0 & 461.17 & 0.0 \\
265.0 & 15.849 & 1.0 & 214.668 & 0.0 & 393.33 & 0.0 \\
266.0 & 7.704 & 1.0 & 156.076 & 0.015 & 336.976 & 0.0 \\
267.0 & 4.226 & 1.0 & 23.258 & 1.0 & 173.847 & 0.001 \\
268.0 & 26.538 & 1.0 & 68.784 & 1.0 & 194.113 & 0.0 \\
269.0 & 63.022 & 1.0 & 332.82 & 0.0 & 500.602 & 0.0 \\
270.0 & 30.372 & 1.0 & 213.945 & 0.0 & 436.388 & 0.0 \\
271.0 & 13.865 & 1.0 & 77.097 & 0.999 & 226.265 & 0.0 \\
272.0 & 58.77 & 1.0 & 517.054 & 0.0 & 579.082 & 0.0 \\
273.0 & 24.328 & 1.0 & 241.735 & 0.0 & 475.864 & 0.0 \\
274.0 & 7.634 & 1.0 & 38.707 & 1.0 & 181.929 & 0.0 \\
275.0 & 13.581 & 1.0 & 147.486 & 0.045 & 345.214 & 0.0 \\
276.0 & 11.827 & 1.0 & 29.386 & 1.0 & 152.761 & 0.023 \\
277.0 & 13.031 & 1.0 & 62.047 & 1.0 & 220.223 & 0.0 \\
278.0 & 10.365 & 1.0 & 132.859 & 0.199 & 302.359 & 0.0 \\
279.0 & 16.127 & 1.0 & 228.845 & 0.0 & 413.122 & 0.0 \\
280.0 & 19.837 & 1.0 & 111.463 & 0.699 & 253.583 & 0.0 \\
281.0 & 6.265 & 1.0 & 126.962 & 0.314 & 257.738 & 0.0 \\
282.0 & 0.241 & 1.0 & 44.24 & 1.0 & 205.68 & 0.0 \\
283.0 & 12.987 & 1.0 & 87.424 & 0.989 & 253.233 & 0.0 \\
284.0 & 25.126 & 1.0 & 53.085 & 1.0 & 182.837 & 0.0 \\
285.0 & 7.722 & 1.0 & 183.448 & 0.0 & 349.299 & 0.0 \\
286.0 & 16.076 & 1.0 & 192.757 & 0.0 & 412.924 & 0.0 \\
287.0 & 49.047 & 1.0 & 362.602 & 0.0 & 546.314 & 0.0 \\
288.0 & 9.24 & 1.0 & 56.195 & 1.0 & 200.021 & 0.0 \\
289.0 & 14.883 & 1.0 & 134.898 & 0.167 & 352.833 & 0.0 \\
290.0 & 10.653 & 1.0 & 306.135 & 0.0 & 545.105 & 0.0 \\
291.0 & 18.341 & 1.0 & 164.004 & 0.005 & 357.345 & 0.0 \\
292.0 & 7.295 & 1.0 & 24.222 & 1.0 & 162.344 & 0.006 \\
293.0 & 20.482 & 1.0 & 135.564 & 0.157 & 297.903 & 0.0 \\
294.0 & 22.394 & 1.0 & 297.0 & 0.0 & 507.476 & 0.0 \\
295.0 & 17.141 & 1.0 & 33.118 & 1.0 & 168.122 & 0.002 \\
296.0 & 16.292 & 1.0 & 96.93 & 0.94 & 231.425 & 0.0 \\
297.0 & 46.826 & 1.0 & 288.356 & 0.0 & 513.537 & 0.0 \\
298.0 & 9.155 & 1.0 & 43.594 & 1.0 & 164.409 & 0.004 \\
299.0 & 6.548 & 1.0 & 16.789 & 1.0 & 165.821 & 0.004 \\
300.0 & 12.573 & 1.0 & 73.084 & 1.0 & 183.424 & 0.0 \\
301.0 & 13.993 & 1.0 & 99.296 & 0.916 & 252.713 & 0.0 \\
302.0 & 24.115 & 1.0 & 232.246 & 0.0 & 348.193 & 0.0 \\
303.0 & 19.542 & 1.0 & 94.719 & 0.957 & 283.896 & 0.0 \\
304.0 & 7.201 & 1.0 & 87.155 & 0.989 & 226.982 & 0.0 \\
305.0 & 15.104 & 1.0 & 213.231 & 0.0 & 392.304 & 0.0 \\
306.0 & 17.707 & 1.0 & 245.732 & 0.0 & 405.259 & 0.0 \\
307.0 & 12.122 & 1.0 & 67.383 & 1.0 & 197.863 & 0.0 \\
308.0 & 48.798 & 1.0 & 390.469 & 0.0 & 527.183 & 0.0 \\
309.0 & 91.851 & 0.974 & 132.103 & 0.212 & 273.432 & 0.0 \\
310.0 & 12.079 & 1.0 & 35.916 & 1.0 & 157.679 & 0.012 \\
312.0 & 46.449 & 1.0 & 90.18 & 0.981 & 238.621 & 0.0 \\
313.0 & 29.506 & 1.0 & 82.872 & 0.996 & 199.859 & 0.0 \\
314.0 & 11.578 & 1.0 & 240.751 & 0.0 & 441.29 & 0.0 \\
315.0 & 28.038 & 1.0 & 49.196 & 1.0 & 169.383 & 0.002 \\
316.0 & 19.462 & 1.0 & 163.854 & 0.005 & 347.544 & 0.0 \\
317.0 & 33.181 & 1.0 & 203.705 & 0.0 & 341.237 & 0.0 \\
318.0 & 26.727 & 1.0 & 273.126 & 0.0 & 484.392 & 0.0 \\
319.0 & 22.109 & 1.0 & 167.369 & 0.003 & 342.045 & 0.0 \\
320.0 & 33.841 & 1.0 & 307.787 & 0.0 & 529.464 & 0.0 \\
321.0 & 7.812 & 1.0 & 70.081 & 1.0 & 247.421 & 0.0 \\
322.0 & 19.73 & 1.0 & 104.177 & 0.848 & 275.152 & 0.0 \\
323.0 & 69.851 & 1.0 & 427.801 & 0.0 & 553.736 & 0.0 \\
324.0 & 62.079 & 1.0 & 246.056 & 0.0 & 474.371 & 0.0 \\
325.0 & 8.426 & 1.0 & 39.38 & 1.0 & 178.004 & 0.0 \\
326.0 & 87.373 & 0.989 & 416.495 & 0.0 & 554.036 & 0.0 \\
327.0 & 60.225 & 1.0 & 401.778 & 0.0 & 594.129 & 0.0 \\
328.0 & 6.036 & 1.0 & 38.218 & 1.0 & 165.726 & 0.004 \\
329.0 & 30.927 & 1.0 & 198.078 & 0.0 & 379.55 & 0.0 \\
330.0 & 12.441 & 1.0 & 29.418 & 1.0 & 158.467 & 0.011 \\
331.0 & 11.892 & 1.0 & 97.908 & 0.931 & 232.279 & 0.0 \\
332.0 & 7.372 & 1.0 & 134.171 & 0.178 & 314.046 & 0.0 \\
333.0 & 9.464 & 1.0 & 40.094 & 1.0 & 168.197 & 0.002 \\
334.0 & 34.678 & 1.0 & 177.828 & 0.0 & 340.089 & 0.0 \\
335.0 & 17.562 & 1.0 & 180.365 & 0.0 & 437.844 & 0.0 \\
336.0 & 34.751 & 1.0 & 283.42 & 0.0 & 420.393 & 0.0 \\
337.0 & 7.388 & 1.0 & 51.622 & 1.0 & 179.153 & 0.0 \\
338.0 & 7.558 & 1.0 & 177.43 & 0.001 & 369.627 & 0.0 \\
339.0 & 6.468 & 1.0 & 18.412 & 1.0 & 158.535 & 0.011 \\
340.0 & 60.323 & 1.0 & 276.247 & 0.0 & 531.049 & 0.0 \\
341.0 & 22.309 & 1.0 & 261.01 & 0.0 & 410.539 & 0.0 \\
342.0 & 6.735 & 1.0 & 198.262 & 0.0 & 385.642 & 0.0 \\
343.0 & 3.997 & 1.0 & 45.52 & 1.0 & 194.711 & 0.0 \\
344.0 & 15.462 & 1.0 & 173.327 & 0.001 & 347.861 & 0.0 \\
345.0 & 4.97 & 1.0 & 162.456 & 0.006 & 309.051 & 0.0 \\
346.0 & 18.127 & 1.0 & 285.618 & 0.0 & 506.617 & 0.0 \\
347.0 & 3.468 & 1.0 & 60.829 & 1.0 & 209.149 & 0.0 \\
348.0 & 29.704 & 1.0 & 151.009 & 0.029 & 296.938 & 0.0 \\
349.0 & 30.889 & 1.0 & 65.662 & 1.0 & 176.772 & 0.001 \\
350.0 & 7.986 & 1.0 & 24.161 & 1.0 & 156.105 & 0.015 \\
351.0 & 8.381 & 1.0 & 167.316 & 0.003 & 320.226 & 0.0 \\
352.0 & 6.881 & 1.0 & 124.535 & 0.37 & 287.507 & 0.0 \\
353.0 & 66.987 & 1.0 & 344.564 & 0.0 & 590.566 & 0.0 \\
354.0 & 16.011 & 1.0 & 216.221 & 0.0 & 364.691 & 0.0 \\
355.0 & 47.382 & 1.0 & 310.992 & 0.0 & 568.263 & 0.0 \\
356.0 & 23.685 & 1.0 & 275.766 & 0.0 & 477.097 & 0.0 \\
357.0 & 10.097 & 1.0 & 72.363 & 1.0 & 235.299 & 0.0 \\
359.0 & 31.797 & 1.0 & 147.22 & 0.046 & 315.634 & 0.0 \\
360.0 & 19.683 & 1.0 & 244.408 & 0.0 & 439.674 & 0.0 \\
361.0 & 6.914 & 1.0 & 35.494 & 1.0 & 173.351 & 0.001 \\
362.0 & 96.571 & 0.943 & 315.967 & 0.0 & 526.681 & 0.0 \\
363.0 & 8.971 & 1.0 & 140.571 & 0.097 & 268.899 & 0.0 \\
364.0 & 29.817 & 1.0 & 306.85 & 0.0 & 608.289 & 0.0 \\
365.0 & 36.033 & 1.0 & 116.502 & 0.573 & 234.041 & 0.0 \\
366.0 & 18.626 & 1.0 & 189.743 & 0.0 & 380.805 & 0.0 \\
367.0 & 77.394 & 0.999 & 445.048 & 0.0 & 574.643 & 0.0 \\
368.0 & 24.242 & 1.0 & 282.767 & 0.0 & 471.912 & 0.0 \\
370.0 & 11.585 & 1.0 & 149.61 & 0.035 & 325.857 & 0.0 \\
371.0 & 12.997 & 1.0 & 150.476 & 0.031 & 355.316 & 0.0 \\
372.0 & 9.967 & 1.0 & 73.858 & 1.0 & 240.909 & 0.0 \\
373.0 & 2.839 & 1.0 & 18.775 & 1.0 & 156.827 & 0.013 \\
374.0 & 21.106 & 1.0 & 88.132 & 0.987 & 222.01 & 0.0 \\
375.0 & 21.664 & 1.0 & 53.217 & 1.0 & 166.045 & 0.003 \\
376.0 & 32.276 & 1.0 & 321.571 & 0.0 & 544.845 & 0.0 \\
377.0 & 2.692 & 1.0 & 20.915 & 1.0 & 154.878 & 0.018 \\
378.0 & 8.674 & 1.0 & 108.206 & 0.772 & 272.277 & 0.0 \\
379.0 & 9.961 & 1.0 & 232.152 & 0.0 & 431.572 & 0.0 \\
380.0 & 15.975 & 1.0 & 106.434 & 0.807 & 265.719 & 0.0 \\
381.0 & 34.322 & 1.0 & 381.526 & 0.0 & 571.546 & 0.0 \\
383.0 & 11.816 & 1.0 & 109.999 & 0.733 & 255.483 & 0.0 \\
384.0 & 2.569 & 1.0 & 24.365 & 1.0 & 196.613 & 0.0 \\
385.0 & 3.167 & 1.0 & 50.792 & 1.0 & 202.307 & 0.0 \\
386.0 & 47.582 & 1.0 & 271.29 & 0.0 & 493.633 & 0.0 \\
387.0 & 26.813 & 1.0 & 321.77 & 0.0 & 560.574 & 0.0 \\
388.0 & 9.891 & 1.0 & 42.131 & 1.0 & 195.11 & 0.0 \\
389.0 & 31.271 & 1.0 & 202.38 & 0.0 & 360.6 & 0.0 \\
391.0 & 14.823 & 1.0 & 54.94 & 1.0 & 187.941 & 0.0 \\
392.0 & 19.4 & 1.0 & 218.097 & 0.0 & 426.132 & 0.0 \\
393.0 & 10.54 & 1.0 & 90.221 & 0.981 & 228.058 & 0.0 \\
394.0 & 0.522 & 1.0 & 54.526 & 1.0 & 224.218 & 0.0 \\
395.0 & 41.788 & 1.0 & 361.936 & 0.0 & 628.46 & 0.0 \\
396.0 & 6.855 & 1.0 & 72.956 & 1.0 & 208.337 & 0.0 \\
397.0 & 10.796 & 1.0 & 310.395 & 0.0 & 527.827 & 0.0 \\
398.0 & 17.522 & 1.0 & 55.46 & 1.0 & 181.073 & 0.0 \\
399.0 & 7.966 & 1.0 & 23.665 & 1.0 & 157.519 & 0.012 \\
400.0 & 8.963 & 1.0 & 184.203 & 0.0 & 384.913 & 0.0 \\

\end{longtable}
\end{document}